\pdfoutput=1
\documentclass[11pt]{article}

\usepackage{graphicx}
\usepackage{amsmath}
\usepackage{amssymb}
\usepackage{dcolumn}
\usepackage{bm}
\usepackage{slashed}
\usepackage{color}
\usepackage{authblk}
\usepackage{ulem}
\usepackage{CJK}

\usepackage{hyperref}
\hypersetup{
	colorlinks=true,
	linkcolor=blue,
	anchorcolor=blue,
	citecolor=blue}

\usepackage{physics}
\usepackage{amsmath}
\usepackage{tikz}
\usepackage{mathdots}
\usepackage{cancel}
\usepackage{color}
\usepackage{array}
\usepackage{multirow}
\usepackage{amssymb}
\usepackage{textcomp, gensymb}
\usepackage{tabularx}
\usepackage{booktabs}
\usetikzlibrary{fadings}
\usetikzlibrary{patterns}
\usetikzlibrary{shadows.blur}
\usetikzlibrary{shapes}

\usepackage{caption}
\usepackage{graphicx}
\usepackage{float} 
\usepackage{subcaption}

\def\L{{\Lambda}}
\def\d{{\delta}}

\def\O{{\Omega}}
\def\e{{\epsilon}}

\def\a{{\alpha}}
\def\b{{\beta}}

\def\f{{\phi}}

\def\G{{\Gamma}}

\def\h{\eta}
\def\p{{\pi}}
\def\P{{\Pi}}
\def\m{{\mu}}
\def\n{{\nu}}
\def\r{{\rho}}
\def\s{{\sigma}}
\def\S{{\Sigma}}

\def\th{{\theta}}

\def\ps{{\psi}}

\def\P{{\Pi}}

\def\({\left(}
\def\){\right)}
\def\[{\left[}
\def\]{\right]}

\newcommand{\lag}{\langle}
\newcommand{\rag}{\rangle}

\newcommand{\pd}{{\partial}}
\newcommand{\dg}{\dagger}

\newcommand{\rel}{\text{rel}}

\date{\today}

\begin{document}

\begin{CJK}{UTF8}{gbsn}

\title{\bf Polarized Dissociation and Spin Alignment of Moving Quarkonium in Quark-Gluon Plasma}

\author[1]{{Zhishun Chen}
\thanks{chenzhsh25@mail2.sysu.edu.cn}}
\affil[1]{School of Physics and Astronomy, Sun Yat-Sen University, Zhuhai 519082, China}
\author[1]{{Shu Lin}
\thanks{linshu8@mail.sysu.edu.cn}}

\maketitle


\begin{abstract}
Recent experiments have found spin alignment of $J/\ps$ with respect to event plane in heavy ion collisions, suggesting a medium effect that is spin dependent. We propose a possible mechanism with polarized dissociation from the motion of $J/\ps$ with respect to the medium.
We calculate polarized dissociation rate for  quarkonium spin triplet state from spin chromomagnetic coupling in the potential non-relativistic QCD framework. This is done for the leading order gluo-dissociation process and next to leading order inelastic Coulomb scattering process. The polarized dissociation rate is expressed as a function of relative velocity between quarkonium and QGP and the quantization axis. Applying the polarized dissociation rate to quarkonium evolution with dissociation effect only in a Bjorken flow, we find the spin $0$ state to dissociate less than the other spin states, leading to positive $\r_{00}-1/3$. Regeneration contribution is expected to give a contribution with the opposite sign.
\end{abstract}

\newpage

\section{Introduction}

It has been predicted based on spin-orbit coupling that the final particles in off-central relativistic heavy ion collisions (HIC) are spin polarized, which includes spin polarization of baryons \cite{Liang:2004ph} and spin alignment of vector meson \cite{Liang:2004xn}. Experimental evidence of the polarization phenomena was first found in global polarization of $\L$ hyperon, with the signal most prominent at low energy collisions \cite{STAR:2017ckg}. The energy dependence of the polarization has been well understood based on spin response to vorticity \cite{Becattini:2013fla,Fang:2016vpj,Li:2017slc}, which characterizes the rotation of quark-gluon plasma (QGP) and its magnitude is known to decrease with increasing collision energy \cite{Jiang:2016woz,Wei:2018zfb}. More recently spin alignment of $\f$ and $J/\ps$ were also found by STAR collaboration \cite{STAR:2022fan} and ALICE collaboration \cite{ALICE:2022dyy} respectively. The vector meson spin states are labeled by $+1$, $-1$ and $0$ with a quantization axis chosen to be perpendicular to the event plane in experiments. The measurement for $\f$ shows that there is significantly more $0$ state than the other two states, with a similar energy dependence to that of $\L$ \cite{STAR:2022fan}, i.e. decreased spin alignment with increasing collision energy. In contrast, the measurement for $J/\ps$ shows a different trend, with less $0$ state and the signal is found at very high energy collisions \cite{ALICE:2022dyy}.

The sharp difference in the experimental results seems to indicate different origins of spin alignment for $\f$ and $J/\ps$. Indeed, $\f$ is believed to be produced from coalescence of strange and anti-strange quarks in QGP \cite{Pal:2002aw}. Different mechanisms have been proposed to understand the spin alignment of $\f$ \cite{Sheng:2022wsy,Sheng:2023urn,Xu:2024kdh,Kumar:2023ghs,Li:2022vmb,Wagner:2022gza,Sheng:2024kgg,Fu:2023qht,Zhao:2024ipr}, see \cite{Chen:2024afy} for a recent review. The production of $J/\ps$ is more complicated: it can be produced from initial hard scattering, which partially dissociate in QGP, from coalescence of charm and anti-charm quarks in QGP, and from feed down of heavier particles. Most $J/\ps$ are produced by the first two processes, which will be referred to as dissociation and regeneration. Unlike the regeneration of $J/\ps$, which is similar to coalescence production of $\f$, the partly dissolved initial $J/\ps$ is a unique component. At high collision energies and low momentum, the regeneration yield is believed to dominate with $J/\ps$ spectrum being nearly thermal \cite{Braun-Munzinger:2000csl}. At intermediate collision energies and large momentum, the dissociation component can be significant \cite{Zhao:2007hh,Zhou:2014kka}.

The dissociation component depends crucially on the dissociation rate in QGP. To have nonvanishing contribution to spin alignment, a spin dependent dissociation rate or polarized dissociation rate is needed. With quarkonium described by non-relativistic QCD \cite{Bodwin:1994jh}, its interaction with gluons in QGP can be decomposed into spin independent chromoelectric dipole coupling and spin dependent chromomagnetic coupling \cite{Brambilla:2004jw}. The latter is suppressed with respect to the former by heavy quark mass. The dissociation rate due to spin independent interaction is well understood for a static quarkonium bound state like $J/\ps$, leading to unpolarized dissociation rate. Spin dependent interaction induced dissociation has been considered in \cite{Chen:2017jje}, obtaining a spin averaged dissociation rate. A quantum kinetic framework for quarkonium with spin dependent interaction has been developed recently \cite{Yang:2024ejk}. Polarized dissociation rate is expected in general for a moving quarkonium in QGP: dissociation from spin-chromomagnetic coupling is proportional to fluctuation of chromomagentic fields seen by quarkonium. The fluctuation of chromomagnetic fields is isotropic in QGP frame, but anisotropic in quarkonium frame, thus polarized dissociation is expected. Since experiments measure spin alignment with momentum average, we will need anisotropic momentum spectrum or flow for QGP.

The paper is organized as follows: in Section~\ref{sec_pnrqcd}, we give a brief review of the potential non-relativistic QCD framework, with an emphasis on spin-dependent interaction; in Section~\ref{sec_sde} we study the polarized dissociation rate. We first relate the spin-chromomagnetic coupling induced dissociation to fluctuation of chromomagnetic fields. We then study contributions to dissociation rate from both gluo-dissociation and inelastic Coulomb scatterings, finding polarized dissociation rate for arbitrary direction of quarkonium velocity and quantization axis; in Section~\ref{sec_appl}, we apply the obtained rate to a phenomenological model with Bjorken flow; Section~\ref{sec_outlook} is devoted to conclusion and outlook.

\section{Potential Non-relativistic QCD in real time formalism}\label{sec_pnrqcd}
The potential Non-Relativistic QCD (pNRQCD) is an effective field theory framework used to describe the dynamics of heavy quarkonium systems \cite{Brambilla:2004jw}. It extends the original Non-Relativistic QCD framework \cite{Bodwin:1994jh} by explicitly integrating out the gluons mediating the binding force between the heavy quarks. Its validity relies on the following hierarchy of scales: $m_Q\gg m_Q v\gg m_Q v^2$, where $m_Q$ is the heavy quark mass, $v$ represents the typical relative velocity of a heavy quark-antiquark pair inside a quarkonium, $m_Q v\sim 1/r$ is then the scale of the typical momentum transfer or inverse radius of the bound state, and $m_Q v^2\sim \e_B$ is the scale of the binding energy. Depending on the hierarchy of $m_Qv$ and QCD scale $\L_{\text{QCD}}$, the quarkonium system can be weakly interacting $m_Qv\gg\L_{\text{QCD}}$ and strongly interacting $m_Qv\ll\L_{\text{QCD}}$ \cite{Brambilla:2004jw}. We shall assume the former and perform perturbative analysis. 

\begin{figure}[!htb]
	\centering
		\includegraphics[width=0.9\textwidth]{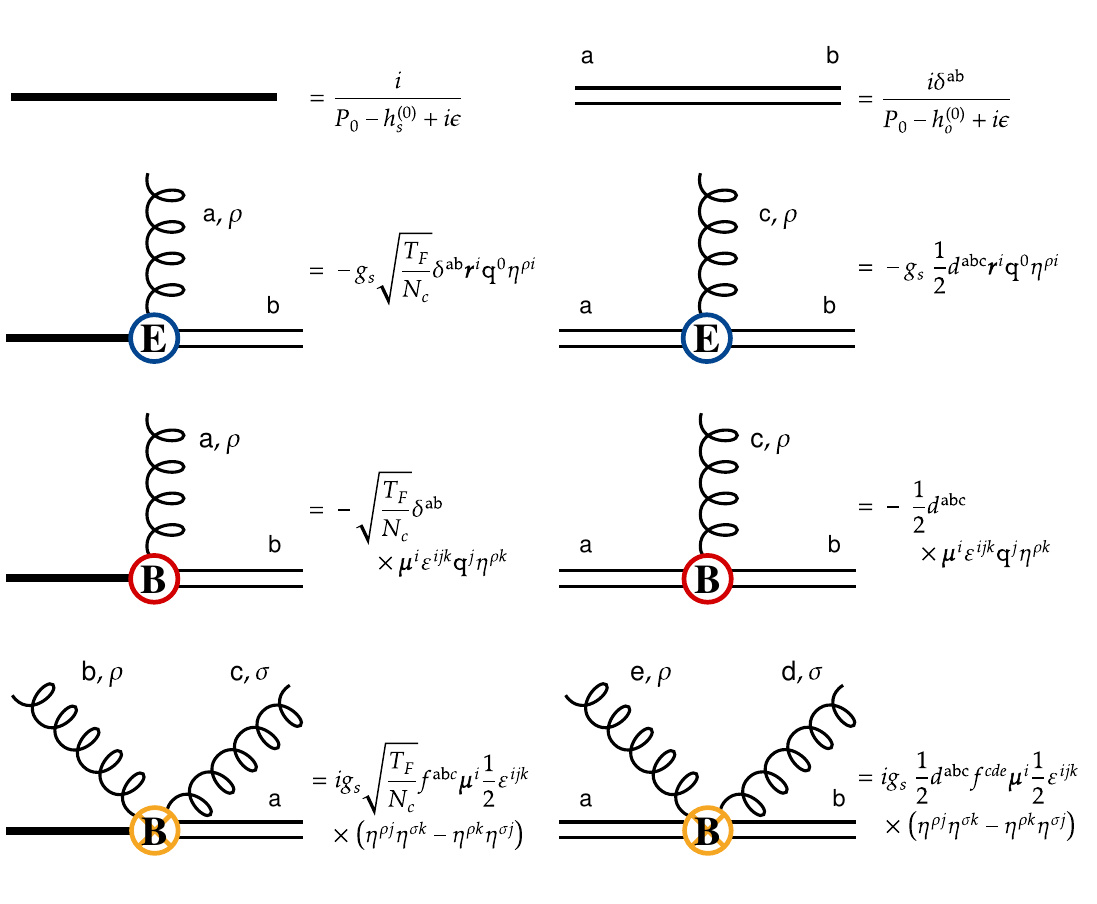}
			    \caption{Feynman rules involving chromomagnetic part in pNRQCD with TAG. The curly line represents gluon. The thick solid line represents quarkonium color singlet and the double solid lines represent quarkonium color octet. $Q^\m$ represents the gluon incoming momentum and $P^\m$ represents the quarkonium momentum. The blue circle with the letter ``E" represents the chromoelectric dipole vertex, while the red circle with the letter ``B" represents the chromomagnetic dipole vertex. chromomagnetic-two-gluon vertex is represented by the orange circle with a cross and the letter ``B". There is no chromoelectric-two-gluon vertex in TAG. Greek and Roman letters correspond to Lorentz and color indices respectively.}
	\label{feynman}
\end{figure}

To describe spin alignment, we need spin dependent interaction, which appears as higher order terms in the multipole expansions of the interaction of quarkonium and gluon \cite{Yan:1980uh}. The corresponding pNRQCD Lagrangian density can be written as
\begin{equation}\label{eq01}
    \begin{split}
    \mathcal{L}_{\text{pNRQCD}}=\mathcal{L}_{\text{light quark}}&+\mathcal{L}_{\text{gluon}}   \\
    +\int & d^3 \bm{r}\, \text{Tr} \Big\{\, \text{S}^{\dag} \[ i\partial_0-h_s \] \text{S}+ \text{O}^{\dag} \[i D_0-h_o \] \text{O} \\
    &+\text{O}^{\dag} g_s\bm{r}\cdot  \bm{E} \text{S}+\text{S}^{\dag} g_s\bm{r}\cdot  \bm{E} \text{O}+\frac{1}{2} \text{O}^{\dag}\{ g_s\bm{r}\cdot  \bm{E},\text{O} \}  \\
    &+\text{O}^{\dag} \bm{\m}\cdot \bm{B} \text{S}+\text{S}^{\dag} \bm{\m}\cdot \bm{B} \text{O}+\frac{1}{2} \text{O}^{\dag}\{ \bm{\m}\cdot \bm{B},\text{O} \}+...\Big\}.
    \end{split}
\end{equation}
Here S$=S \bm{1}_c/\sqrt{N_c}$ and O$=O^a T^a /\sqrt{T_F}$ are the normalized color singlet and color octet fields, respectively. The trace is over spin and color indices. $\bm{E}$ and $\bm{B}$ are the chromoelectric and chromomagnetic fields. 
$g_s\bm{r}$ and $g_s(\bm{\s}_Q-\bm{\s}_{\overline{Q}})/(2m_Q)\equiv\bm{\m}$ represent chromoelectric and chromomagnetic dipoles respectively, with $g_s$ being the strong coupling constant.
The last two lines of \eqref{eq01} correspond to the spin independent chromoelectric dipole and spin dependent chromomagnetic dipole interactions.
We will take leading-order Hamiltonians for the color singlet and color octet:
\begin{equation}\label{eq02}
    \begin{split}
    h_s^{(0)}&=\frac{(i\overline{\nabla}_{\bm{R}})^2}{4m_Q}+\frac{(i\overline{\nabla}_{\bm{r}})^2}{m_Q}-C_F \frac{\a_s}{r},   \\
    h_o^{(0)}&=\frac{(i\overline{\nabla}_{\bm{R}})^2}{4m_Q}+\frac{(i\overline{\nabla}_{\bm{r}})^2}{m_Q}+\frac{1}{2N_c} \frac{\a_s}{r},
    \end{split}
\end{equation}
where $R$ and $r$ represent the center-of-mass coordinate and the relative coordinate respectively. The covariant derivative for the octet field is defined as $D_0 \text{O}=\partial_0 \text{O}-i g_s \[A_0,\text{O}\]$. We will use Temporal Axial Gauge (TAG) $A_0^a=0$ so that $D_0=\pd_0$. The Feynman rules we used are summarized in Fig.~\ref{feynman}.

We will study the dissociation of quarkonium bound state in finite temperature QGP using the real time formalism in $ra$-basis \cite{Ghiglieri:2020dpq}. Below we briefly discuss modifications of Feynman rules in the $ra$-basis using singlet field. Similar modifications apply to octet field. The fields in $ra$-basis are defined as
\begin{align}
	S_r=\frac{S_1+S_2}{2},\quad S_a=S_1-S_2,
\end{align}
with $S_{1,2}$ being fields on Schwinger-Keldysh contour. The Feynman rules in Fig.~\ref{feynman} correspond to all field being $S_1$. In real time formalism in the $ra$-basis, the fields appearing in propagators and vertices can be either $S_r$ or $S_a$. We shall consider dissociation of probe quarkonium in QGP, for which the following vacuum propagators will be used
\begin{align}\label{singlet_prop}
&S_{ra}(P)=\frac{i}{p_0-h_s^{(0)}+i\e}, \nonumber\\
&S_{ar}(P)=\frac{i}{p_0-h_s^{(0)}-i\e}, \nonumber\\
&S_{rr}(P)=\p\d(p_0-h_s^{(0)}).
\end{align}
The vertices in $ra$-basis can be worked out easily as\footnote{Here we use $ra$ labels on $\bm{E}$ instead of on $\bm{A}$. This is justified as $\bm{E}$ is always linear in $\bm{A}$ in TAG. The situation of $\bm{B}$ is complicated by the presence of terms quadratic in $\bm{A}$, for which $ra$ labels should be placed on $\bm{A}$.}
\begin{align}
&O_1^\dg g_s\bm{r}\cdot\bm{E}_1S_1-O_2^\dg g_s\bm{r}\cdot\bm{E}_2S_2\nonumber\\
=&O_r^\dg g_s\bm{r}\cdot\bm{E}_rS_a+O_a^\dg g_s\bm{r}\cdot\bm{E}_rS_r+O_a^\dg g_s\bm{r}\cdot\bm{E}_rS_r+\frac{1}{4}O_a^\dg g_s\bm{r}\cdot\bm{E}_aS_a.
\end{align}
Note that the vertices always contain odd number of $a$ fields.
In our case, only vertices with one $a$ field are needed. We will label $ra$ fields explicitly when necessary. Since the color structure of vertices are the same as the vacuum counterpart, the color indices will be suppressed to avoid conflict with $ra$ labels.

\section{Polarized Dissociation}\label{sec_sde}

The dissociation of a static quarkonium bound state in QGP has been studied at leading order (LO) in the coupling, in which the singlet dissociates into octet by absorbing a gluon in the medium. This is the gluo-dissociation mechanism \cite{Bhanot:1979vb,Peskin:1979va,Song:2005yd,Chen:2017jje}. It has then been realized the LO cross section is suppressed by phase space for small binding energy \cite{Grandchamp:2001pf,Grandchamp:2002wp}. The NLO process comes from one of the quarkonium constituents undergoes inelastic scattering with light quarks and gluons in the QGP. The suppression in coupling constant is compensated by enhancement in the phase space, leading to a significant contribution phenomenologically, see also \cite{Chen:2018dqg,Zhao:2024gxt}. Both LO and NLO contribution have been studied in the pNRQCD framework based on chromoelectric dipole interaction \cite{Brambilla:2011sg,Brambilla:2013dpa,Brambilla:2010vq,Yao:2018sgn}. 
The resulting dissociation rates are spin independent.

Now we consider spin dependent contribution to dissociation rate from the chromomagnetic dipole interaction. The spin averaged dissociation rate for a static quarkonium bound state has already been obtained in \cite{Chen:2017jje}. Below we first give a general account of the spin dependence of dissociation rate. With application to $J/\ps$ in mind, we consider singlet and octet with the following spin and orbital angular momentum
\begin{equation}\label{eq9}
\begin{split}
\text{color singlet}&:\,S=1,\,L=0 \\
\text{color octet}&:\,S=0,\,L=0.
\end{split}
\end{equation}
For a chosen quantization axis $\bm{l}$, the spin triplet states are given by: $\ket{\uparrow \uparrow}$, $\ket{\downarrow \downarrow}$, $\ket{(\uparrow\downarrow +\downarrow\uparrow)/\sqrt{2}}$ and the spin singlet state is given by $\ket{(\uparrow\downarrow -\downarrow\uparrow)/\sqrt{2}}$. Then it is straightforward to work out
\begin{equation}\label{eq10}
\begin{split}
\abs{ \frac{1}{\sqrt{2}}\,\bra{\uparrow\downarrow -\downarrow\uparrow}\, \(\frac{\bm{\sigma}_c}{2}-\frac{\bm{\sigma}_{\overline{c}}}{2}\)\cdot \bm{B} \,\ket{\uparrow \uparrow}}^2&=\frac{B_{l\perp}^2}{2} \\
\abs{ \frac{1}{\sqrt{2}}\,\bra{\uparrow\downarrow -\downarrow\uparrow}\,  \(\frac{\bm{\sigma}_c}{2}-\frac{\bm{\sigma}_{\overline{c}}}{2}\)\cdot \bm{B} \,\ket{\downarrow \downarrow}}^2&=\frac{B_{l\perp}^2}{2} \\
\abs{ \frac{1}{2}\,\bra{\uparrow\downarrow -\downarrow\uparrow}\,  \(\frac{\bm{\sigma}_c}{2}-\frac{\bm{\sigma}_{\overline{c}}}{2}\)\cdot \bm{B} \,\ket{\uparrow\downarrow +\downarrow\uparrow}}^2&=B_l^2, \\
\end{split}
\end{equation}
where $B_l=\bm{B}\cdot\hat{\bm{l}}$ and $B_{l\perp}^2=B^2-B_l^2$. We have suppressed the color indices in \eqref{eq10}. The chromomagnetic field can come from a real gluon in LO gluo-dissociation or a virtual gluon in NLO inelastic scattering. Clearly in isotropic QGP, we have isotropic fluctuation of chromomagnetic field $\lag B_l^2\rag=\lag B_{l\perp}^2\rag/2$. Thus the dissociation is still spin independent. The situation is different when the quarkonium moves in the QGP. The aim of the remaining part of this section is to find out the splitting in the dissociation rates for different spin states.

The dissociation rate for a specific spin state is readable from the resummed propagator satisfying the following equation
\begin{align}
S^{\text{resum}}_{ra}(P)^{-1}=S_{ra}(P)^{-1}-\S_{ar}.
\end{align}
Using the bare propagator in \eqref{singlet_prop}, the above is easily solved as
\begin{align}
S_{ra}^{\text{resum}}=\frac{i}{p_0-h_s^{(0)}-i\S_{ar}}.
\end{align}
It follows that the dissociation rate is given by\footnote{The factor of 2 arises because the width $\G$ describes the decay rate of the probability $\sim|S|^2$, while $\text{Re}[\S_{ar}]$ describes the decay rate of the wave function $S$. Since the probability density decays twice as fast as the amplitude, the factor of 2 appears in the relationship.}
\begin{align}
\G=-2\text{Re}[\S_{ar}].
\end{align}
While the bare propagator is degenerate for different spin states, the self-energy can be spin dependent as argued above, leading to spin dependent dissociation rate.

Before proceeding to the actual calculations, we comment on an additional complication due to the presence of medium: in the presence of finite temperature QGP, two more scales enter: temperature $T$ and gluon thermal mass (to be defined shortly) $m_{\text{g}}\sim g_sT$. We will consider $T\gg\L_{QCD}$ so that $g_s\ll1$ ($m_{\text{g}}\ll T$)\footnote{The coupling among QGP constituents can be different from the coupling between quarkonium constituents in principle. We choose both of them to be weak.}. We further require $m_Qv\gg T$ and $m_{\text{g}}\sim \e_B$\footnote{$\e_B$ is determined by Coulombic potential between the quarkonium constituents. We treat it as a free parameter as it describes the structure of quarkonium, which does not interfere with medium property characterized by $m_g$ etc.}. While not essential for the analysis, these conditions lead to significant simplifications.

    
\subsection{Self-energy for a static quarkonium}

\begin{figure}[!htb]
    \centering
    \includegraphics[width=0.9\textwidth]{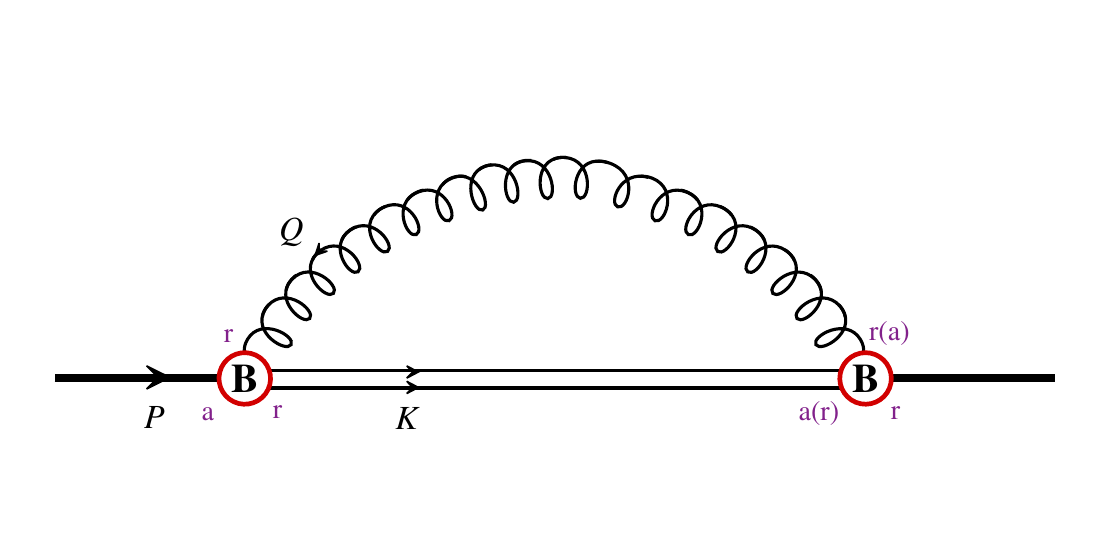}
    \caption{One-loop self-energy diagram for quarkonium color singlet. The vertices correspond to chromomagnetic dipole interaction. Two $ra$ labelings are possible, with only one being medium dependent. \protect}
    \label{fig:one-loop}
\end{figure}

We shall first consider contribution to dissociation rate from chromomagnetic coupling for a static quarkonium bound state in QGP and then generalize to a moving quarkonium bound state.
We start with one-loop diagrams depicted in Fig.~\ref{fig:one-loop}.
They give rise to the following contribution to color singlet self-energy
\begin{align}\label{LO_rep}
    &\S_{ar}(P)=\frac{T_F}{N_c}(N_c^2-1)\int_Q D_{rr}^{mn}(Q)O_{ra}(P+Q)\lag S|\m_i|O\rag\lag O|\m_j|S\rag\e^{ikm}q_k\e^{jln}(-q_l),\\
    &\S_{ar}(P)=\frac{T_F}{N_c}(N_c^2-1)\int_Q D_{ar}^{mn}(Q)O_{rr}(P+Q)\lag S|\m_i|O\rag\lag O|\m_j|S\rag\e^{ikm}q_k\e^{jln}(-q_l),\nonumber
\end{align}
where $\int_Q\equiv\int \frac{d^4Q}{(2\p)^4}$. The color indices have been suppressed, with a factor of $(N_c^2-1)$ inserted from the color sum. $O$ and $D$ denote propagators for octet and gluons respectively. Note that since we take quarkonium as a probe, $O_{rr}$ is medium independent, see \eqref{singlet_prop}. It is clear then the second line corresponds to a medium independent contribution to dissociation rate, which we do not consider. The dissociation rate from the first line is given by
\begin{align}\label{Gamma_LO}
    \G&=-2\text{Re}[\S_{ar}(P)]=2\frac{T_F}{N_c}(N_c^2-1)\int_Q D_{rr}^{mn}(Q)\text{Re}[O_{ra}(P+Q)]\lag S|\m_i|O\rag\lag O|\m_j|S\rag\e^{ikm}q_k\e^{jln}q_l,\nonumber\\
    &=2\frac{T_F}{N_c}(N_c^2-1)\int_Q D_{rr}^{mn}(Q)\pi\d(p_0+q_0-h_o^{(0)})\lag S|\m_i|O\rag\lag O|\m_j|S\rag\e^{ikm}q_k\e^{jln}q_l.
\end{align}
We have assumed that all other factors except $O_{ra}$ are real, which we now justify. The explicit form of $D_{rr}$ is given by
\begin{align}\label{Drr_free}
    D_{rr}^{mn}(Q)=P_T^{mn}(Q)2\pi\e(q_0)\d(Q^2)\(\frac{1}{2}+f(q_0)\),
\end{align}
with $P_T^{mn}=\d_{mn}-\hat{q}^m\hat{q}^n$ being the transverse projector. It is clearly real. We will also drop the medium independent factor $1/2$.
The transition amplitude between the singlet and octet states split into orbital and spin parts as
\begin{align}\label{LS_split}
    \lag S|\m_i|O\rag\lag O|\m_j|S\rag=\lvert\lag S,L=0|O,L=0\rag\rvert^2\lag S=1|\m_i|S=0\rag\lag S=0|\m_j|S=1\rag.
\end{align}
Clearly, the orbital part contributes a real factor $\lvert\lag S,L=0|O,L=0\rag\rvert^2$. The spin part and the remaining factors are calculated as
\begin{align}\label{contracts}
    &\lag S=1|\m_i|S=0\rag\lag S=0|\m_j|S=1\rag\e^{ikm}q_k\e^{jln}q_l P_T^{mn}\nonumber\\
    =&q^2\lag S=1|\m_i|S=0\rag\lag S=0|\m_j|S=1\rag(\d_{ij}-\hat{q}^i\hat{q}^j).
\end{align}
For $0$-state in the triplet, we have from \eqref{eq10}
\begin{align}\label{spin0}
    q^2\lag S=1,S_l=0|\m_i|S=0\rag\lag S=0|\m_j|S=1,S_l=0\rag(\d_{ij}-\hat{q}^i\hat{q}^j)=\frac{g_s^2}{m_Q^2}\(q^2-(\bm{q}\cdot\bm{l})^2\).
\end{align}
The spin summed counterpart is given by
\begin{align}\label{spinsum}
    \sum_{s=0,\pm1}q^2\lag S=1,S_l=s|\m_i|S=0\rag\lag S=0|\m_j|S=1,S_l=s\rag(\d_{ij}-\hat{q}^i\hat{q}^j)=\frac{g_s^2}{m_Q^2}2q^2.
\end{align}
We see that the spin part is also real, which justifies taking the real part on $D_{ra}$ before. To proceed, we note that for on-shell singlet state, we have $p_0=h_s^{(0)}=2m_Q-\e_B$. For the color octet state, we may decompose the relative motion by plane wave basis $h_o^{(0)}=2m_Q+\frac{p_\rel^2}{m_Q}+\frac{q^2}{4m_Q}$, with $p_\rel$ being relative momentum of the pair\footnote{This amounts to ignoring the final state interaction between the quark pair in the octet.}. The last term is the kinetic energy from motion of the center of mass. Ignoring the last term for now, we obtain the spin summed LO dissociation rate from the chromomagnetic dipole interaction as
\begin{align}\label{plane_wave}
    \sum_{s=0,\pm1}\G_s&=2\frac{g_s^2}{m_Q^2}\frac{T_F}{N_c}(N_c^2-1)\int_Q 2\pi\d(Q^2)f(q_0)\pi\d(p_0+q_0-h_o^{(0)})\abs{\lag S,L=0|O,L=0\rag}^2 2q^2\nonumber\\
    &=2\frac{g_s^2}{m_Q^2}\frac{T_F}{N_c}(N_c^2-1)\int\frac{d^3 q}{(2\p)^3 2q}f(q)\p\frac{m_Q p_{rel}}{(2\p)^2}\abs{\lag S,L=0|p_\rel\rag}^2\vert_{p_\rel=\sqrt{m_Q(q-\e_B)}}2q^2.
\end{align}
The Dirac delta function gives $p_\rel=\sqrt{(q_0-\e_B)m_Q}\simeq \sqrt{q_0 m_Q}$. Since $q_0\sim T\ll p_\rel$ by our assumtpion, we can indeed ignore the kinetic energy from the center of mass motion. \eqref{plane_wave} is in agreement with \cite{Chen:2017jje}. 
The dissociation rate for spin $0$ state is easily obtained by noting that the integrand is isotropic in $\bm{q}$, so that we may substitute in \eqref{spin0} $\hat{q}^i\hat{q}^j\to1/3\,\d_{ij}$ to find that the dissociation rate for the spin $0$ state is $1/3$ of the spin summed counterpart. This is in line with our general discussion earlier that isotropic fluctuation of chromomagnetic fields lead to no spin alignment.

\begin{figure}[!htb]
    \centering
    \includegraphics[width=0.9\textwidth]{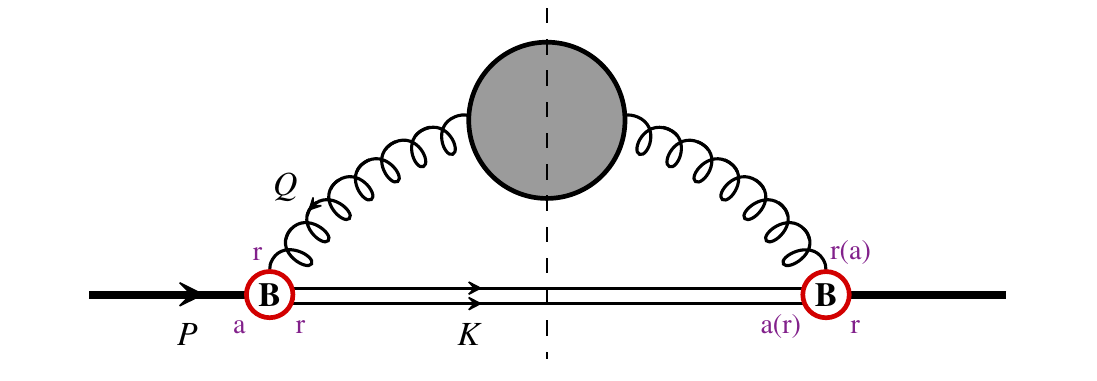}
    \caption{Effective one-loop self-energy diagram for quarkonium color singlet. The vertices correspond to chromomagnetic dipole interaction. The gluon propagator with blob stands for resummed propagator. Two $ra$ labelings are possible, with only one being medium dependent.\protect}
    \label{fig:resummed_loop}
\end{figure}
Next, we turn to the effective one-loop diagrams depicted in Fig.~\ref{fig:resummed_loop}. In this case, we have similar representation as \eqref{LO_rep} for this contribution
\begin{align}\label{Gamma_NLO}
\G&=2\frac{T_F}{N_c}(N_c^2-1)\int_Q  D_{rr}^{mn}(Q)\pi\d(p_0+q_0-h_o^{(0)})\lag S|\m_i|O\rag\lag O|\m_j|S\rag\e^{ikm}q_k\e^{jln}q_l,
\end{align}
with $D_{rr}^{mn}$ being resummed gluon propagator. Its explicit form in TAG is given by \cite{Bellac:2011kqa}
\begin{align}\label{Drr_nu}
    &D_{rr}^{mn}=\r^{mn}(Q)\(\frac{1}{2}+f(q_0)\),\nonumber\\
    &\r^{mn}(Q)=\r_T(Q) P_T^{mn}+\r_L(Q)\frac{q^2}{q_0^2}\hat{q}^m\hat{q}^n,
\end{align}
where $\r_T=2\text{Im}[\frac{-1}{Q^2-G}]$ and $\r_L=2\text{Im}[\frac{-1}{Q^2-F}\frac{Q^2}{q^2}]$ are transverse and longitudinal spectral functions. $F$ and $G$ are longitudinal and transverse components of self-energy given by
\begin{align}\label{FG}
     F&=-\frac{2m_{\text{g}}^2 Q^2}{q^2}\( 1-\frac{q_0}{q} Q_0\( \frac{q_0}{q} \) \),\nonumber  \\
     G&=\frac{1}{2}\( 2m_{\text{g}}^2-F \),
\end{align}
in the hard thermal loop (HTL) approximation\footnote{The self-energy is gauge invariant in the HTL regime.}. $m_{\text{g}}^2=\frac{1}{6}g_s^2T^2(C_A+\frac{N_f}{2})$ is the gluon thermal mass. The spectral functions contain pole and cut contributions. In the limit $Q\gg m_{\text{g}}$, the spectral functions are dominated by pole contributions with the transverse pole reducing to the free theory spectral function \eqref{Drr_free} while the longitudinal pole decouples. The pole contribution is also suppressed by phase space, similar to the case of LO calculation. The cut contribution corresponds to the inelastic Coulomb scattering when expanded to first order in $F$ and $G$ \cite{Brambilla:2013dpa}. When $Q\sim m_{\text{g}}$, the Coulomb scattering contains infrared (IR) divergence, which is partially screened by self-energy and the remaining divergence is cutoff by $\e_B$, resulting in logarithmically enhanced contribution of the form $\ln\frac{T}{\e_B}$ and $\ln\frac{T}{m_{\text{g}}}$ to the dissociation rate \cite{Brambilla:2010vq}.
Below we shall focus on the following cut contribution (and loosely refer to it as NLO contribution)
\begin{align}\label{Gamma_cut}
    \G&=2\frac{T_F}{N_c}(N_c^2-1)\int_Q \r_{\text{cut}}^{mn}\(\frac{1}{2}+f(q_0)\)\pi\d(p_0+q_0-h_o^{(0)})\lag S|\m_i|O\rag\lag O|\m_j|S\rag\e^{ikm}q_k\e^{jln}q_l,
\end{align}
with
$\r_{\text{cut}}^{mn}=2\p(P_T^{mn}\b_T+\hat{q}^m\hat{q}^n\b_L)$. The explicit expressions of $\b_{T/L}$ are given by
\begin{align}\label{beta}
    &\b_L=\frac{m_{\text{g}}^2 x\th (1-x^2)}{\left[q^2+2m_{\text{g}}^2\left(1-\frac{x}{2}\ln \abs{\frac{x+1}{x-1}}\right)\right]^2+\p^2 m_{\text{g}}^4 x^2},\nonumber   \\
    &\b_T=\frac{m_{\text{g}}^2x(1-x^2)\th(1-x^2)/2}{\left(q^2(x^2-1)-m_{\text{g}}^2\left[x^2+\frac{x(1-x^2)}{2}\ln\abs{\frac{x+1}{x-1}}\right]\right)^2+\p^2m_{\text{g}}^4x^2(1-x^2)^2/4},
\end{align}
with $x=q_0/q$.
Like in the LO case, the integrand is isotropic in $\bm{q}$, so the argument we have used for LO case still works, leading to an unpolarized dissociation rate with no spin alignment.

\begin{figure}[!htb]
  \centering
  \subfloat[\label{TAG1}]{
    \includegraphics[width=0.3\textwidth]{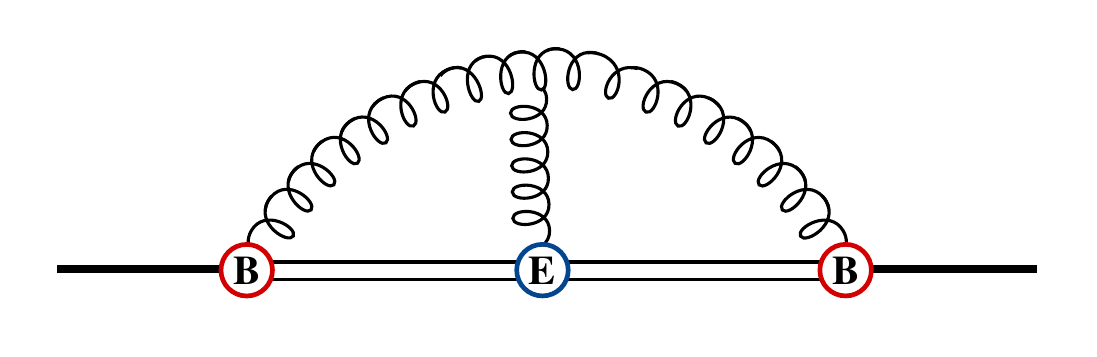}
  }
  \subfloat[\label{TAG2}]
  {
    \includegraphics[width=0.3\textwidth]{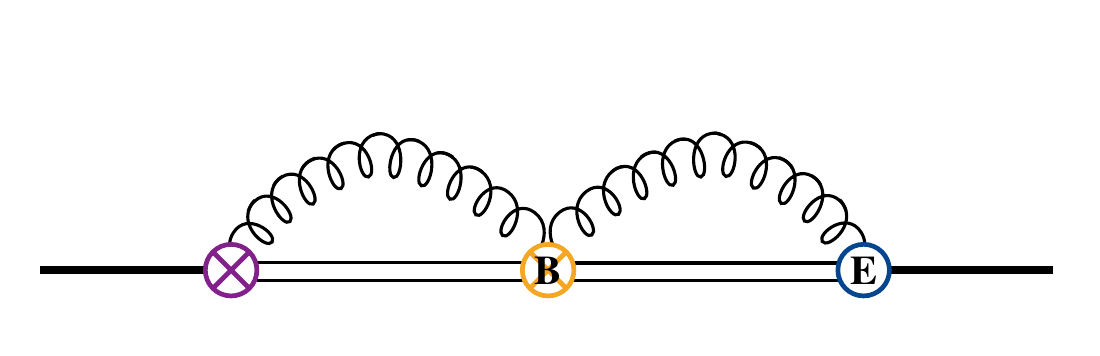}
  }
  \subfloat[\label{TAG3}]
  {
    \includegraphics[width=0.3\textwidth]{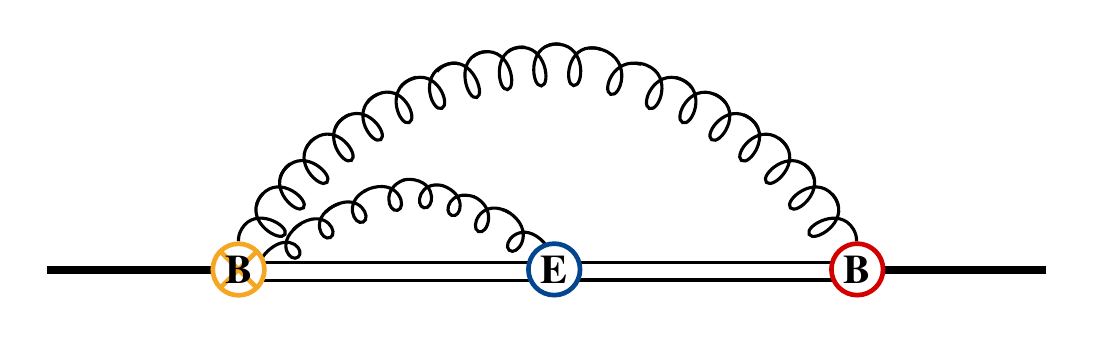}
  }
  
  \subfloat[\label{TAG4}]{
    \includegraphics[width=0.45\textwidth]{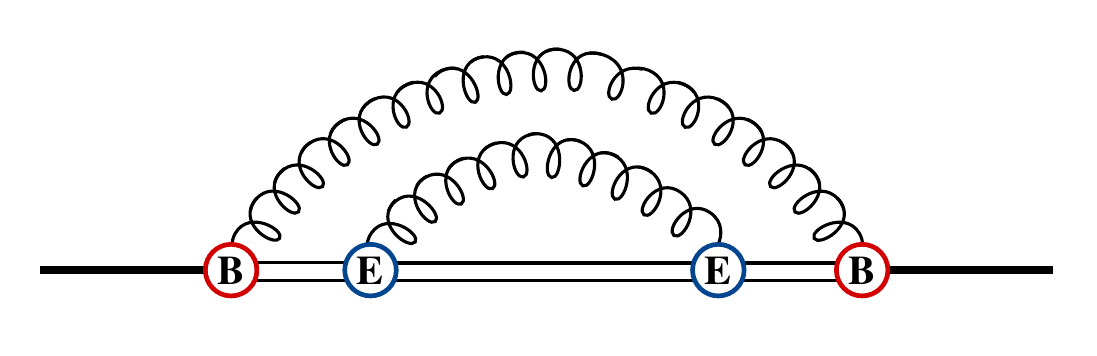}
  }
  \subfloat[\label{TAG5}]
  {
    \includegraphics[width=0.45\textwidth]{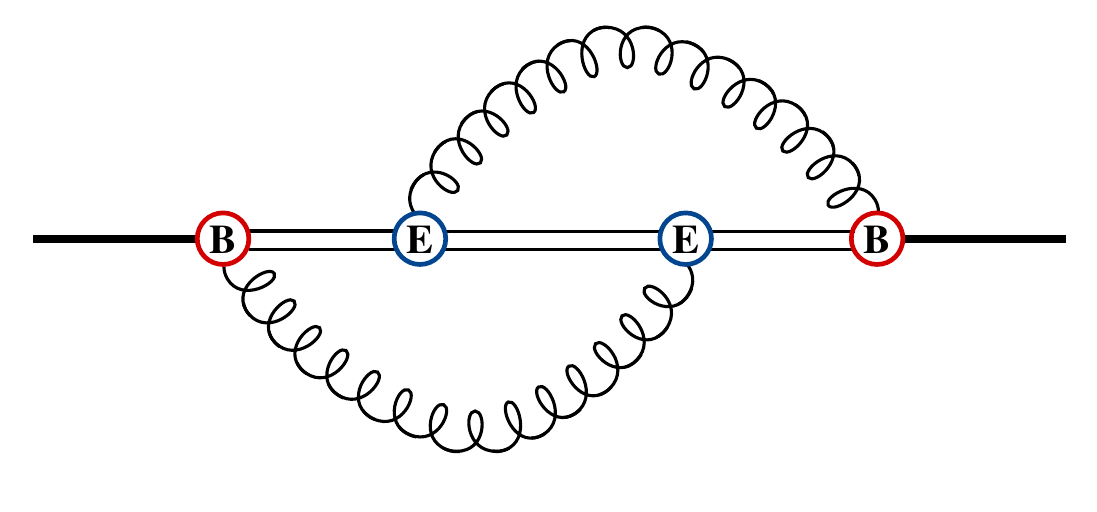}
  }

  \subfloat[\label{TAG6}]
  {
    \includegraphics[width=0.45\textwidth]{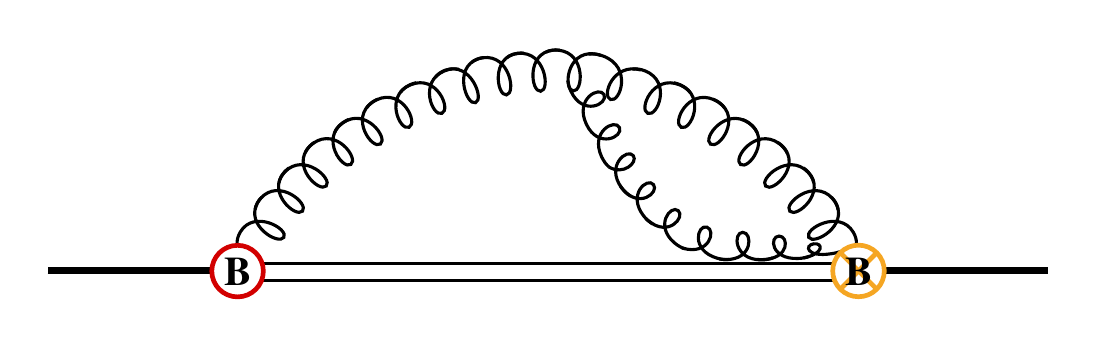}
  }
  \subfloat[\label{TAG7}]{
    \includegraphics[width=0.45\textwidth]{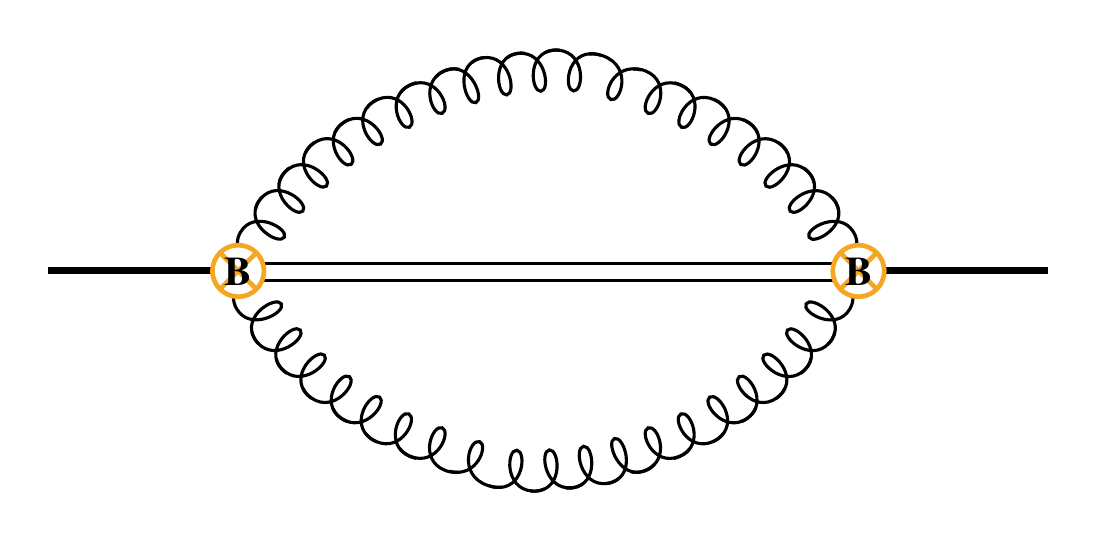}
  }
  \caption{Diagrams involving chromoelectric vertices and non-Abelian chromomagnetic vertices for quarkonium self-energy on NLO level. The purple circle with a cross can be chromoelectric or chromomagnetic dipoles. They diagrams give rise to interference correction to gluo-dissociation or Compton scattering contribution. They either give parametrically suppressed contributions or not leading logarithmically enhanced. See the text for more discussions. 
  \protect}
  \label{TAG}
\end{figure}

Fig.~\ref{fig:resummed_loop} contains only a subset of two-loop diagrams. The other diagrams are collected in Fig.~\ref{TAG}. These diagrams can be cut in different ways to give rise to: interference correction to gluo-dissociation from diagrams (a)-(f); Compton like scattering from diagrams (a), (c)-(g)\footnote{Diagram in Fig.~\ref{fig:resummed_loop} with gluon self-energy from gluon loop also corresponds to Compton scattering in the t-channel in this terminology.}. Although the diagrams contain the same powers of $g_s$ as those for Coulomb scattering, most of them are suppressed by extra powers of $rq_0\sim \frac{T}{m_Qv}\ll 1$ by our assumption. The exceptions are diagram (b) when the cross vertex is taken to be chromoelectric interaction and diagram (g). The former case vanishes for the following reason: the virtual color octet states sandwiching the chromomagnetic vertex inherit the spin states of the color singlet from chromoelectric vertices, thus are spin triplet states. Their matrix elements vanish on the chromomagnetic vertex. The latter case does give nonvanishing contribution to polarized dissociation rate. This contribution is, however, not IR enhanced like in the Coulomb scattering case. Below we shall restrict ourselves to the leading logarithmic contributions and leave more comprehensive studies for future work.

\subsection{Self-energy for a moving quarkonium}

Now we consider dissociation of a moving quarkonium bound state. Since the spin states are defined in quarkonium rest frame, it is natural to think of the problem in the same frame, in which static quarkonium bound state dissociates in moving QGP. The chromoelectromagnetic fields seen by the quarkonium is a boosted version of the isotropic fields in the QGP frame, which are expected to be anisotropic. Through chromomagnetic coupling, we expect a polarized dissociation rate and nonvanishing spin alignment.

It is convenient to proceed in covariant form. We denote the quarkonium and QGP frames respectively by $n^\m$ and $u^\m$. An ambiguity arises in the definition of TAG, which can be either $A^a\cdot n=0$ or $A^a\cdot u=0$. We choose the former, which keeps the Feynman rules in Fig.~\ref{feynman} unchanged. The only change we need is to replace the gluon propagators in \eqref{Gamma_LO} and \eqref{Gamma_NLO} by the following moving ones. Note that the gluons are in equilibrium so the fluctuation-dissipation theorem holds in general
\begin{align}\label{FDT}
    D_{rr}^{\m\n}=2\text{Re}[D_{ra}^{\m\n}]\(\frac{1}{2}+f(Q\cdot u)\).
\end{align}
The replacement for the free gluon propagator (indicated by the superscript $(0)$) is straightforward 
\begin{align}\label{Drr0_n}
    D_{rr}^{\m\n(0)}&=P_T^{\m\n}(n)2\p\e(Q\cdot n)\d(Q^2)\(\frac{1}{2}+f(Q\cdot u)\).
\end{align}
Here the transverse projector defined as $P_T^{\m \n}(n)=n^\m n^\n-\h^{\m\n}-\hat{q}_n^\m\hat{q}_n^\n$ with $\hat{q}_n^\m=(-n^\m n^\n+\h^{\m\n})Q_\n/(Q\cdot n)^2-Q^2)^{1/2}$ being covariant generalization of the unit vector $\hat{\bm{q}}$ in quarkonium frame. The factor $1/2$ will be dropped as medium independent contribution.

The replacement for the resummed gluon propagator is tricky. $D_{ra}^{\m\n}$ satisfies the following resummation equation
\begin{align}\label{resum_eq}
    D_{ra}^{\m\n}=D_{ra}^{\m\n(0)}+D_{ra}^{\m\a(0)}\P_{ar,\a\b}D_{ra}^{\b\n}.
\end{align}
The free retarded gluon propagator in TAG is given by
\begin{align}\label{eq25n}
    D^{(0)\,\m \n}_{ra}=\frac{i}{Q^2} P_T^{\m \n}(n)+\frac{i}{Q^2}\frac{Q^2}{(Q\cdot n)^2 }\hat{q}_n^\m \hat{q}_n^\n,
\end{align}
On the other hand, the gluons are in equilibrium in QGP frame with the following self-energy
\begin{align}
    \P_{ar}^{\m \n}(u)=P_T^{\m \n}(u)\, G(u)+P_L^{\m \n}(u)\, F(u),
\end{align}
where the transverse projector $P_T^{\m\n}(u)$ are defined similarly as $P_T^{\m\n}(n)$ but with different frame vector. $P_L^{\m\n}(u)=\h^{\m\n}-Q^\m Q^\n/Q^2-P_T^{\m\n}(u)$ is the longitudinal projector. The components of self-energy are covariant generalization of \eqref{FG} as $G(u)=G(q_0=Q\cdot u,q=((Q\cdot u)^2-Q^2)^{1/2})$ and $F(u)=F(q_0=Q\cdot u,q=((Q\cdot u)^2-Q^2)^{1/2})$. For $n=u$, it is not difficult to arrive at \eqref{Drr_nu}. For $n\ne u$, the solution to the resummation equation is complicated.

To simplify the analysis, we assume the quarknonium is slow-moving in QGP and expand in the velocity. Working in the quarkonium frame, we parameterize $u^\m=((1+\d\bm{v}^2)^{1/2},\d\bm{v})$. We already know from the previous subsection that there is no spin alignment when $\d\bm{v}=0$. It follows that the spin alignment arises at least from $\mathcal{O}(\d\bm{v}^2)$\footnote{The naive contribution to spin alignment of the form $\mathcal{O}(\d\bm{v}\cdot\hat{\bm{l}})$ is excluded because the spin alignment is invariant under flipping of quantization axis.}. We shall solve for the resummed gluon propagator up to $\mathcal{O}(\d\bm{v}^2)$. \eqref{resum_eq} is solved as $D_{ra}=( I-D^{(0)}_{ra} \P_{ar} )^{-1} D^{(0)}_{ra}$ with all quantities being matrices. To perform the expansion, we split the gluon self-energy as
\begin{align}\label{Pi_split}
    \P_{ar,\m\n}(u)=\P_{ar,\m\n}(u)-\P_{ar,\m\n}(n)+\P_{ar,\m\n}(n)\equiv \d\P_{ar,\m\n}+\P_{ar,\m\n}(n).
\end{align}
The matrix inversion is performed using
\begin{align}
    (M-\d M)^{-1}=M^{-1}+M^{-1} \d M M^{-1}+\cdots.
\end{align}
to arrive at
\begin{align}\label{dDra}
    \d D_{ra}^{\m\n}=D_{ra}^{\m\a}(n)\d\P^{ar}_{\a\b}D_{ra}^{\b\n}(n)+\cdots,
\end{align}
with $D_{ra}(n)$ in the above being the resummed propagator given by
\begin{align}
    D_{ra}^{\m \n}(n)=\frac{i}{Q^2-G(n)} P_T^{\m \n}(n)+\frac{i}{Q^2-F(n)}\frac{Q^2}{(Q\cdot n)^2 }\hat{q}_n^\m \hat{q}_n^\n.
\end{align}
The terms containing one more pair of $\d\P^{ar}$ and $D_{ra}$ is suppressed by $m_{\text{g}}^2/Q^2$. As will be clear soon, the dominant contribution comes from the phase space with $q_0,q\gg m_{\text{g}}$, allowing us to keep the leading term only.
By the fluctuation-dissipation theorem \eqref{FDT}, we find
\begin{align}
    \d D_{rr}^{\m\n}&=2\text{Re}[ D_{ra}^{\m\n}(u)]\(\frac{1}{2}+f(Q\cdot u)\)-2\text{Re}[ D_{ra}^{\m\n}(n)]\(\frac{1}{2}+f(Q\cdot n)\).
\end{align}
Clearly, by \eqref{dDra} $\d D_{rr}^{\m\n}$ is nonvanishing for spatial components only in quarkonium frame.
$2\text{Re}[\d D_{ra}^{\m\n}]$ can be interpreted as a correction to the spectral density, although strictly speaking it is only a consequence of our splitting of the self-energy.

\subsection{LO dissociation rate}\label{LO}
We start with the LO dissociation for a moving quarkonium bound state. The LO dissociation rate written in quarkonium frame reads
\begin{align}\label{Gamma_LO_moving}
\G&=2\frac{T_F}{N_c}(N_c^2-1)\int_Q D_{rr}^{mn(0)}(n)\pi\d(p_0+q_0-h_o^{(0)})\lag S|\m_i|O\rag\lag O|\m_j|S\rag\e^{ikm}q_k\e^{jln}q_l,
\end{align}
with $D_{rr}^{mn(0)}$ being spatial components of $D_{rr}^{\m\n(0)}$. Plugging \eqref{contracts} and \eqref{Drr0_n} into \eqref{Gamma_LO_moving}, we obtain
\begin{align}\label{parameterize}
\G&=2\frac{T_F}{N_c}(N_c^2-1)\int_Q2\p^2\frac{1}{2q}\d(q_0-q)\d(p_0+q_0-h_o^{(0)})f(Q\cdot u)\lag S|\m_i|O\rag\lag O|\m_j|S\rag q^2(\d^{ij}-\hat{q}^i\hat{q}^j)\nonumber\\
&\equiv\G_{ij}\lag S=1|\m_i|S=0\rag\lag S=0|\m_j|S=1\rag\frac{m_Q^2}{g_s^2}.
\end{align}
We used \eqref{LS_split} and dropped the $1/2$ next to $f(Q\cdot u)$ because it does not contribute to the spin dependence of the dissociation rate. A normalization factor $m_Q^2/g_s^2$ is introduced for convenience. 
The spin part of the transition amplitude is obtained from \eqref{eq10} as
\begin{align}\label{spin_part}
&\text{spin 0-state}:\;\lag S=1|\m_i|S=0\rag\lag S=0|\m_j|S=1\rag=\frac{g_s^2l_il_j}{m_Q^2},\nonumber\\
&\text{spin summed}:\;\lag S=1|\m_i|S=0\rag\lag S=0|\m_j|S=1\rag=\frac{g_s^2\d_{ij}}{m_Q^2}.
\end{align}
We then express the splitting of dissociation rate as
\begin{align}
\overline{\G}-\G_{s=0}=\G_{ij}\(\frac{1}{3}\d_{ij}-{l}_i{l}_j\),
\end{align}
where $\overline{\G}=1/3\,\sum_{s=0,\pm1}\G_s$. 
The explicit form of $\G_{ij}$ is determined as follows
\begin{align}
\G_{ij}=2\frac{T_F}{N_c}(N_c^2-1)\int_Q 2\p^2\frac{1}{2q}\d(q_0-q)\d(p_0+q_0-h_o^{(0)})f(Q\cdot u)\lag S|\m_i|O\rag\lag O|\m_j|S\rag q^2(\d^{ij}-\hat{q}^i\hat{q}^j).
\end{align}
The orbital part of the transition amplitudes is evaluated as
\begin{align}\label{amplitude}
|\lag S,L=0|O,L=0\rag|^2=\int\frac{d^3p_\rel}{(2\p)^3}\frac{64\p a^3}{(1+a^2p_\rel^2)^3},
\end{align}
with $a=(m_Q\e_B)^{-1/2}$ being the Bohr radius for the bound state. As in the static case, since $q\lesssim T\ll m_Qv$, the kinetic energy of the center of mass motion is suppressed with respect to the counterpart from relative motion of quark pair, so that $\d(p_0+q_0-h_o^{(0)})\simeq\d(-\e_B+q_0-\frac{p_\rel^2}{m_Q})$. It follows that
\begin{align}\label{prel_int}
\int\frac{d^3p_\rel}{(2\p)^3}\frac{a^3}{(1+a^2p_\rel^2)^3}\d(-\e_B+q_0-\frac{p_\rel^2}{m_Q})=\frac{1}{(2\p)^2}\frac{m_Qa^3p_\rel}{(1+a^2p_\rel^2)^3}.
\end{align}
evaluated at $p_\rel=((q_0-\e_B)m_Q)^{1/2}$.

To simplify the tensor integral of $\G^{ij}$, we parameterize $\G_{ij}=c_1\d_{ij}+c_2\d\hat{v}^i\d\hat{v}^j$ by rotational invariance. Only $c_2$ is relevant for splitting of dissociation rate, which can be easily determined as
\begin{align}\label{c2}
c_2&=-\frac{1}{2}\G^{ij}\d_{ij}+\frac{3}{2}\G^{ij}\d\hat{v}_i\d\hat{v}_j,
\end{align}
which is a scalar integral depending on magnitude of $\d v$ only. It is calculated numerically. The dissociation rate in terms of $c_2$ reads
\begin{align}\label{LOG}
\overline{\G}-\G_{s=0}&=c_2 \(\frac{1}{3}-\(\d\hat{\bm{v}}\cdot\hat{\bm{l}}\)^2\).
\end{align}
The coefficient $c_2$ is a function of $T$, magnitude of relative velocity $\d v$, $\e_B$ and $m_Q$. The spin dependence is entirely contained in the factor $1/3-(\d\hat{\bm{v}}\cdot\hat{\bm{l}})^2$. 
\begin{figure}[!htb]
	\centering
		\includegraphics[width=0.9\textwidth]{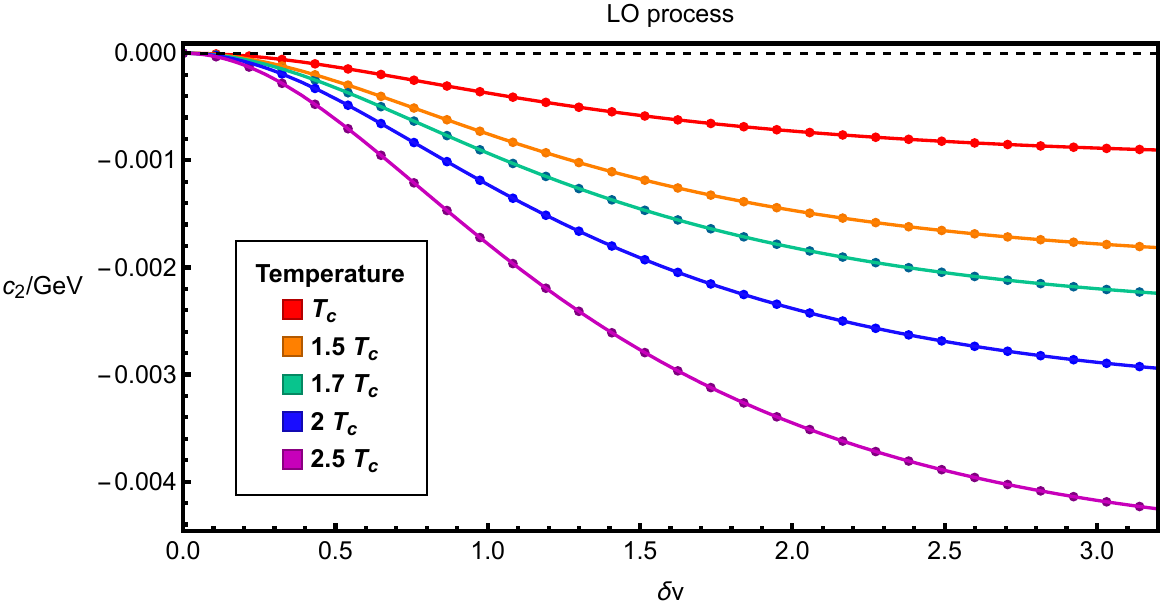}
			    \caption{Numerical results of magnitude of splitting in dissociation rate $c_2$ versus $\d v$ at different temperatures. $c_2$ is the key parameter related to the spin alignment, which together with some constants and angle $\d \hat{\bm{v}}\cdot\hat{\bm{l}}$ determines the splitting of the spin dependent dissociation rate. Here we take $\a_s=0.3$ and binding energy $\e_B=0.052$GeV.   \protect}
	\label{c2LO}
\end{figure}
Taking $J/\psi$ as an example, in the numerical integration, we take $\a_s=0.3$, $T_c=150$MeV and $m_c=1.3$GeV for charm quark. Using 1S wave function, $J/\ps$ has a definite mass $M=2m_c-\epsilon_B$ and binding energy $\epsilon_B=m_c C_F^2 \alpha_s^2/4=0.052$GeV. The dependence of $c_2$ on $\d v$ at different temperatures is shown in Fig.~\ref{c2LO}. We can see that $c_2$ is always negative, then the sign of the splitting of dissociation rate depends on the angle between $\d \bm{v}$ and $\hat{\bm{l}}$: when $\abs{\d\hat{\bm{v}}\cdot\hat{\bm{l}}}<1/\sqrt{3}$, $\overline{\G}<\G_{s=0}$, otherwise the result changes sign.

\subsection{NLO dissociation rate}\label{NLO}
We turn to the NLO dissociation, with the following expression for the rate
\begin{align}\label{Gamma_NLO_moving}
\G&=2\frac{T_F}{N_c}(N_c^2-1)\int_Q \d D_{rr}^{mn}(Q)\pi\d(p_0+q_0-h_o^{(0)})\lag S|\m_i|O\rag\lag O|\m_j|S\rag\e^{ikm}q_k\e^{jln}q_l,    
\end{align}
with $\d D_{rr}^{mn}$ being spatial components of $\d D_{rr}^{\m\n}$ from \eqref{Drr0_n}. It can be further simplified by noting that the chromomagnetic vertices are transverse to $Q$, so only the transverse components of $\d D_{ra}^{mn}$ survive upon contraction. We may expand $\d D_{ra}$ as
\begin{align}
    &\d D_{ra}^{mn(1)}=P_T^{ma}(n)\frac{i}{Q^2-G(n)}\d\P_{ab}^{(1)}P_T^{bn}(n)\frac{i}{Q^2-G(n)},\nonumber\\
    &\d D_{ra}^{mn(2)}=P_T^{ma}(n)\frac{i}{Q^2-G(n)}\d\P_{ab}^{(2)}P_T^{bn}(n)\frac{i}{Q^2-G(n)}.
\end{align}
$\d\P_{ab}$ are spatial components of $\d\P_{ar,\m\n}$ defined in \eqref{Pi_split}. The subscript $ar$ is suppressed to avoid conflict with spatial indices. The superscripts $(1)$ and $(2)$ indicate the corresponding quantities expanded to $\mathcal{O}(\d \bm{v})$ and $\mathcal{O}(\d \bm{v}^2)$. To obtain explicit expansion of $\d\P$, we denote $\d P^T_{ab}=P^T_{ab}(u)-P^T_{ab}(n)$, $\d P^L_{ab}=P^L_{ab}(u)-P^L_{ab}(n)$, $\d G=G(u)-G(n)$ and $\d F=F(u)-F(n)$ and rewrite $\d\P_{ab}$ as
\begin{align}\label{dPi_split}
    i\d\P_{ab}&=\d P^T_{ab}G(n)+\d P^L_{ab}F(n)+P_{ab}^T\d G+P^L_{ab}\d F+\d P^T_{ab}\d G+\d P^L_{ab}\d F,
\end{align}
Note that $\d\P_{ab}$ is to be contracted with transverse projectors on both sides, we can equivalently replace $\d P^T_{ab}\to -\frac{Q^2}{(Q\cdot u)^2-Q^2}u_a u_b$, which is already $\mathcal{O}(\d \bm{v}^2)$. Meanwhile $\d P^T+\d P^L=0$. This leads to the following expansions of $\d\P$:
\begin{align}
    i\d\P_{ab}^{(1)}&=P_{ab}^T(n)\d G^{(1)}+P_{ab}^L(n)\d F^{(1)},\nonumber\\
    i\d\P_{ab}^{(2)}&=\d P_{ab}^T(G(n)-F(n))+P_{ab}^T(n)\d G^{(2)}+P_{ab}^L(n)\d G^{(2)}.
\end{align}
We then obtain the following correction to spectral density
\begin{align}
    \d \r^{mn}&=2\text{Re}[\d D_{ra}^{mn}]\nonumber\\
    \d \r^{mn(1)}&=-2\text{Im}\[\frac{1}{(Q^2-G)^2}\d G^{(1)}\]P_{mn}^T(n),\nonumber\\
    \d \r^{mn(2)}&=-2\text{Im}\[\(\frac{1}{Q^2-G(n)}\)^2(3G(n)-2m_{\text{g}}^2)\]\frac{-Q^2u_a u_b}{(Q\cdot n)^2-Q^2}P_T^{ma}P_T^{bn}\nonumber\\
    &-2\text{Im}\[\(\frac{1}{Q^2-G(n)}\)^2\d G^{(2)}\]P_{mn}^T(n).
\end{align}
We have used $F/2+G=m_{\text{g}}^2$ to simplify the expressions. Again the superscripts $(1)$ and $(2)$ indicate the corresponding quantities expanded to $\mathcal{O}(\d \bm{v})$ and $\mathcal{O}(\d \bm{v}^2)$.
With same logic as before, we keep the imaginary parts from those of $G$ and $F$ etc only, i.e. without including pole contributions. This can be viewed as a correction to the cut contribution from the motion of QGP.  
The corresponding correction to $\d D_{rr}^{mn}$ reads 
\begin{align}
    \d D_{rr}^{mn}=\d\r^{mn}\(\frac{1}{2}+f(Q\cdot n)\)+\r^{mn}(n)\d f+\d\r^{mn}\d f,
\end{align}
with $\d f=f(Q\cdot u)-f(Q\cdot n)$. We only need it expanded to $\mathcal{O}(\d \bm{v}^2)$, which is explicitly given by
\begin{align}\label{Drr2}
    &\d D_{rr}^{mn(2)}=-2\text{Im}\[\(\frac{1}{Q^2-G(n)}\)^2(3G(n)-2m_{\text{g}}^2)\]\frac{-Q^2u_a u_b}{(Q\cdot n)^2-Q^2}P_T^{ma}P_T^{bn}\(\frac{1}{2}+f(Q\cdot n)\)\nonumber\\
    &-2\text{Im}\[\(\frac{1}{Q^2-G(n)}\)^2\d G^{(2)}\]P_{mn}^T(n)\(\frac{1}{2}+f(Q\cdot n)\)\nonumber\\
    &-2\text{Im}\[\frac{1}{Q^2-G(n)}\]P_{mn}^T(n)\d f^{(2)}-2\text{Im}\[\frac{1}{(Q^2-G(n))^2}\d G^{(1)}\]P_{mn}^T(n)\d f^{(1)}.
\end{align}
The first line arises from correction to projectors, and the last two lines arise from correction to spectral function $\d G$ and correction to distribution function $\d f$.

To consider the transition amplitude, we still evaluate the orbital and spin parts. For the orbital part, we again use plane wave for the evaluation of the orbital part to have \eqref{amplitude} and \eqref{prel_int}. For the spin part, we need the following contractions
\begin{align}\label{contractions}
    &\lag S=1|\m_i|S=0\rag\lag S=0|\m_j|S=1\rag\e^{ikm}q_k\e^{jln}q_l P_T^{mn}(\hat{\bm{q}}\cdot\d\bm{v})^2,\nonumber\\
    &\lag S=1|\m_i|S=0\rag\lag S=0|\m_j|S=1\rag\e^{ikm}q_k\e^{jln}q_l P_T^{ma}u_a u_b P_T^{bn}.
\end{align}
We have included a factor of $(\hat{\bm{q}}\cdot\d\bm{v})^2$ in the first line for the following reason: the corresponding projector $P_T^{mn}$ from the last two lines of \eqref{Drr2} is accompanied by one of the three factors $\d G^{(2)}$, $\d f^{(2)}$ and $\d G^{(1)}\d f^{(1)}$, which all gives the factor $(\hat{\bm{q}}\cdot \d \bm{v})^2$ when expanded to $\mathcal{O}(\d \bm{v}^2)$.

Note that both $P_T^{mn}$ and $P_T^{ma}u_a u_b P_T^{bn}$ are transverse to $\bm{q}_n$. This allows us to replace the combination $\e^{ikm}\e^{jln}$ by its transversely projected one as
\begin{align}
    \e^{ikm}\e^{jln}q_k q_l\to q^2(\d_{ij}\d_{mn}-\d_{in}\d_{jm})-\d_{mn}q_iq_j.
\end{align}
We collect the contractions needed up on using the replacement above
\begin{align}\label{contractions_exp}
    &\e^{ikm}q_k\e^{jln}q_l P_T^{mn}(\hat{\bm{q}}\cdot\d\bm{v})^2=(\bm{q}\cdot\d\bm{v})^2\(\d^{ij}-\hat{q}^i\hat{q}^j\),\nonumber\\
    &\e^{ikm}q_k\e^{jln}q_l P_T^{ma}u_a u_b P_T^{bn}=q^2\(\hat{q}^i\hat{q}^j\(\hat{\bm{q}}\cdot\d\bm{v}\)^2-\hat{\bm{q}}^{\{i}\d\bm{v}^{j\}}\(\hat{\bm{q}}\cdot\d\bm{v}\)+\d v^i \d v^j\).
\end{align}

Before performing the phase space integral, we argue that the dominant contribution arise from the region with $\e_B\sim m_{\text{g}}\ll q_0(q)\ll T$ if we are concerned with the logarithmically enhanced contributions. This can be in the form of either $\ln(T/\e_B)$ or $\ln(T/m_{\text{g}})$. The former originates from IR divergence cut off by binding energy, which implies $\e_B\ll q_0$. The other condition $q_0\ll T$ is needed to provide the additional Bose-Einstein enhancement factor $1/2+f(Q\cdot u)\simeq (\b q_0)^{-1}$. The latter originates from collinear divergence cut off by thermal mass. In fact, the restriction on phase space $q_0(q)\ll T$ is also satisfied. This follows from the factor \eqref{prel_int} that diminishes for $a p_\rel\gg1$, which occurs for $q_0\gg \e_B$ but can still be well below the scale of $T$. Thus the approximation $1/2+f(Q\cdot u)\simeq (\b q_0)^{-1}$ is also justified for the collinearly enhanced case.

\begin{figure}[!htb]
	\centering
		\includegraphics[width=0.9\textwidth]{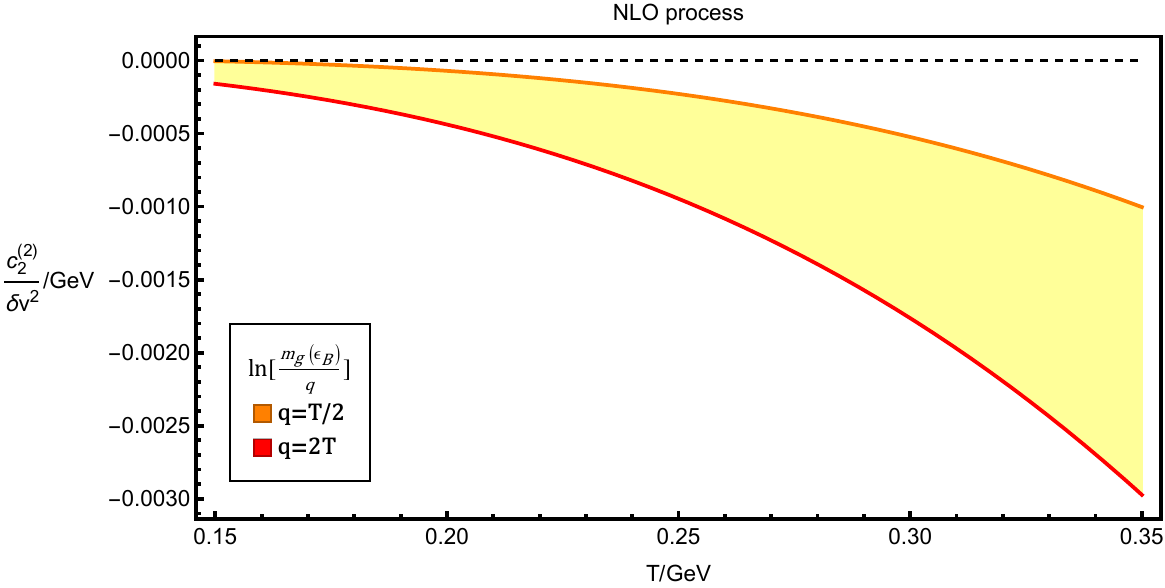}
			    \caption{Numerical results of $c_2^{(2)}$ for $J/\ps$, shows the dependence of $c_2^{(2)}/\d v^2$ on temperature $T$, the superscript $(2)$ means the results take to $\mathcal{O}(\d \bm{v}^2)$ order. Theory uncertainty is estimated by varying the $q$-integration upper bound from $T/2$ to $2T$ (see the inset). We take $\a_s=0.3$, then binding energy $\e_B=0.052$ GeV and gluon thermal mass $m_\text{g}\simeq1.5853 T$.  \protect}
	\label{c2NLO}
\end{figure}

Now we are ready to perform the phase space integral with $\e_B\sim m_{\text{g}}\ll q_0(q)\ll T$. The identification of the phase space allows for significantly simplification of the integration. The details are reserved in appendix~\ref{sec_app_A}. As shown in \eqref{parameterize} in LO calculation, we can still define $\G^{ij\,(2)}=c_1^{(2)}\d^{ij}+c_2^{(2)}\d \hat{v}^i \d \hat{v}^j$ to obtain the key parameter $c_2^{(2)}$ relevant for the splitting of dissociation rate. Plugging \eqref{contractions_exp} \eqref{int0} \eqref{int1} \eqref{int2} into \eqref{Drr2} and \eqref{Gamma_NLO_moving}, we obtain
\begin{align}\label{NLO_Gamma}
    \G=&-2\frac{T_F}{N_c}(N_c^2-1)\frac{1}{16\p^4}\int_{-1}^{1}d\(\hat{\bm{q}}\cdot\d\bm{v}\)\lag S=1|\m_i|S=0\rag\lag S=0|\m_j|S=1\rag\nonumber  \\
    &\Bigg(\frac{3\p^2m_{\text{g}}^2}{16\b}\ln\[\frac{\e_B}{T}\]\(\hat{q}^i\hat{q}^j\(\hat{\bm{q}}\cdot\d\bm{v}\)^2-\hat{\bm{q}}^{\{i}\d\bm{v}^{j\}}\(\hat{\bm{q}}\cdot\d\bm{v}\)+\d v^i \d v^j\)\nonumber  \\
    &\,\,+\frac{3 \pi ^2 m_{\text{g}}^2}{32 \beta}\left(3\ln\[\frac{m_{\text{g}}}{T}\] +\ln \[\frac{\e_B}{T}\]\right)\,\(\hat{\bm{q}}\cdot\d\bm{v}\)^2\(\d^{ij}-\hat{q}^i\hat{q}^j\)\Bigg)\nonumber \\
    \equiv&\G^{ij\,(2)}\lag S=1|\m_i|S=0\rag\lag S=0|\m_j|S=1\rag\frac{m_Q^2}{g_s^2}.
\end{align}
And the parameter $c_2^{(2)}$ and the splitting of dissociation rate have the same form as \eqref{c2}\eqref{LOG}
\begin{align}\label{c22}
c_2^{(2)}=&-\frac{1}{2}\G^{ij\,(2)}\d_{ij}+\frac{3}{2}\G^{ij\,(2)}\d\hat{v}_i\d\hat{v}_j,\nonumber  \\
\overline{\G}^{(2)}-\G_{s=0}^{(2)}=&c_2^{(2)} \(\frac{1}{3}-\(\d\hat{\bm{v}}\cdot\hat{\bm{l}}\)^2\).
\end{align}
For the NLO process we focus on the logarithmically enhanced contributions. Simulating in $J/\ps$'s rest frame, the corresponding $c_2^{(2)}/\d v^2$ is shown in Fig.~\ref{c2NLO}, based on the fact that we expand to the order $\mathcal{O}(\d v^2)$, this quantity will only be a function of $T$. The logarithm have forms such as $\ln\frac{\e_B}{T}$ and $\ln\frac{m_\text{g}}{T}$, which are the results from taking the integration upper bound at $q=T$. We will vary the upper bound from $q=T/2$ to $q=2T$ as estimate for theory uncertainty.

\section{Spin alignment in Bjorken flow}\label{sec_appl}

To calculate the effect of polarized dissociation on spin alignment, we need to implement the dissociation rate in a transport model. We study quarkonium evolution in QGP with the following Boltzmann equation for quarkonium
\begin{equation}\label{eq1}
    P^{\mu }{\partial_\mu }f_i\(x,P\)=-C_i\(x,P\)f_i\(x,P\),
\end{equation}
where $i=+1,-1,0$ denotes different spin states, $P^{\mu}$ represents the 4-momentum of quarkonium, $u^{\mu}$ is the flow velocity. In \eqref{eq1} we have kept the dissociation contribution and neglected the regeneration contribution. $C$ is the dissociation coefficient arising from both chromoelectric and chromomagnetic dipole transition. The coefficient relates to dissociation rate as follows: $C_i=(P\cdot n)\G_i$.
With the distribution function $f_i$ solved for the different spin states, the spin alignment can then be written as
\begin{equation}\label{eq2}
    \rho _{00}-\frac{1}{3}=\frac{f^0}{f^0+f^-+f^+}-\frac{1}{3}.
\end{equation}
For illustration purpose, we take the QGP flow to be Bjorken flow, which allows us to obtain analytic expressions. In Bjorken flow, a convenient representation is the Milner coordinate, using the following parameterizations: $t=\tau\,\text{cosh}(\eta)$, $z=\tau\,\text{sinh}(\eta)$ with proper time $\tau=\sqrt{t^2-z^2}$ and space-rapidity $\eta=1/2\,\text{ln}[(t+z)/(t-z)]$. Then the flow velocity and 4-momentum of quarkonium can be written as $u^{\mu}=(\text{cosh}(\eta),\bm{0},\text{sinh}(\eta))$ and $P^{\mu}=\(M_T\,\text{cosh}\(Y\),\bm{p_T},M_T\,\text{sinh}\(Y\)\)$ with momentum-rapidity $Y=1/2\,\text{ln}[(E+p_z)/(E-p_z)]$ and transverse mass $M_T=\sqrt{M^2+\bm{{p_T}}^2}$. In such a representation, \eqref{eq1} becomes
\begin{equation}\label{eq3}
    \(\partial_{\tau }+\frac{1}{\tau}\text{tanh}\(Y-\eta\) \partial _{\eta }\)f_i(\tau,\eta,Y,\bm{p_T})=-\frac{C_i}{P\cdot u}f_i(\tau,\eta,Y,\bm{p_T})\equiv-\frac{1}{\tau _R}f_i(\tau,\eta,Y,\bm{p_T}),
\end{equation}
the solution is essentially worked out in \cite{Baym:1984np,Jaiswal:2021uvv}. 
Let us make a further assumption introduced in \cite{Zhu:2004nw}: All quarkonia are produced at $t=z=0$. We then have $Y=\eta$ hold throughout the evolution. With this assumption, distribution function can be written as 
\begin{align}\label{ansatz}
f_i\(\tau ,\eta ,Y,\bm{p_T}\)=(\tau _0/\tau)\, \widetilde{f}_i\(\tau ,Y,\bm{p_T}\)\delta (\eta -Y).    
\end{align}
The physical reason is that dissociation does not change rapidity and momentum distribution. With \eqref{ansatz}, \eqref{eq3} has a special solution
\begin{equation}\label{eq4}
    \widetilde{f}_i\(\tau ,Y,\bm{p_T}\)=e^{-\int_{\tau _0}^{\tau } d\tau ^{\prime} \frac{1}{\tau _R}}\widetilde{f}_{i}\(\tau_0 ,Y,\bm{p_T}\).
\end{equation}
Since $1/\tau_R=(C^E+C{^B_i})/(P\cdot u)$ is a small quantity, the expansion of the exponential factor can be performed to simplify the solution as 
		\begin{equation}\label{eq5}
			\widetilde{f}_i\(\tau ,Y,\bm{p_T}\)=e^{-\int_{\tau _0}^{\tau } d\tau ^{\prime} \, \frac{C^E}{P\cdot u}}\(1-\int_{\tau _0}^{\tau } d\tau ^{\prime} \, \frac{C{^B_i}}{P\cdot u}+\mathcal{O}\(\(\int_{\tau _0}^{\tau } d\tau ^{\prime} \, \frac{C{^B_i}}{P\cdot u}\)^2\)\)\widetilde{f}_{i}\(\tau_0 ,Y,\bm{p_T}\).
		\end{equation}
		Assuming initial distribution of quarkonia are spin independent, we can
	plug \eqref{eq5} into \eqref{eq2} to find the chromoelectric part and initial distribution function are nicely canceled out to give
		\begin{equation}\label{eq6}
			\rho _{00}-\frac{1}{3}\simeq -\frac{1}{3} \int_{\tau _0}^{\tau } d\tau ^{\prime} \, \frac{C{^B_0}}{P\cdot u}+\frac{1}{3} \int_{\tau _0}^{\tau } d\tau ^{\prime} \, \frac{\overline{C}^B}{P\cdot u}.
		\end{equation}
We will then consider LO and NLO dissociation rates separately in \eqref{eq6}.
%
The splitting of LO and NLO dissociation rate are given by \eqref{LOG} and \eqref{c22} respectively, which give rise to the spin alignment as
 \begin{align}
	&\rho^{\text{LO}} _{00}-\frac{1}{3}= \frac{1}{3} \int_{\tau_0}^{\tau} d\tau ^{\prime} \, \frac{P\cdot n}{P\cdot u}c_2 \(\frac{1}{3}-\(\d\hat{\bm{v}}\cdot\hat{\bm{l}}\)^2\),\nonumber\\
    &\rho^{\text{NLO}} _{00}-\frac{1}{3}=\frac{1}{3} \int_{\tau_0}^{\tau} d\tau ^{\prime} \, \frac{P\cdot n}{P\cdot u}c_2^{(2)} \(\frac{1}{3}-\(\d\hat{\bm{v}}\cdot\hat{\bm{l}}\)^2\).
\end{align}

\begin{figure}[!htb]
  \centering
  \includegraphics[width=0.9\textwidth]{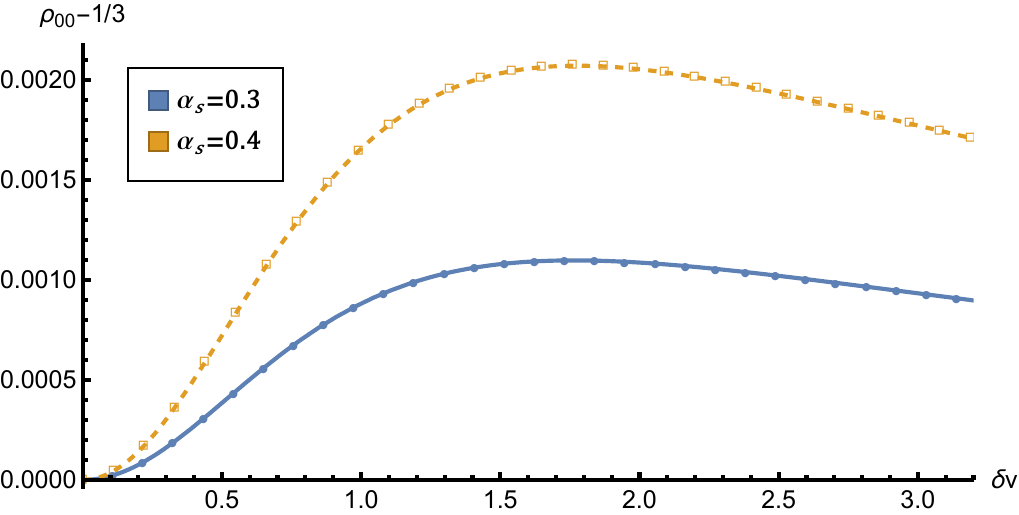}
  \caption{Numerical fitting result of spin alignment of the LO process calculated in quarkonium's rest frame. The blue filled-circle and orange empty-square plots are the calculated data, while blue solid curve and orange dashed curve are the fitting function represent strong coupling constant $\alpha_s$ takes 0.3 and 0.4, respectively. Then binding energy $\e_B=0.052$GeV and $0.0924$GeV in these two cases. \protect}
  \label{LOresult}
\end{figure}

\begin{figure}[ht]
    \centering
    \includegraphics[width=0.9\textwidth]{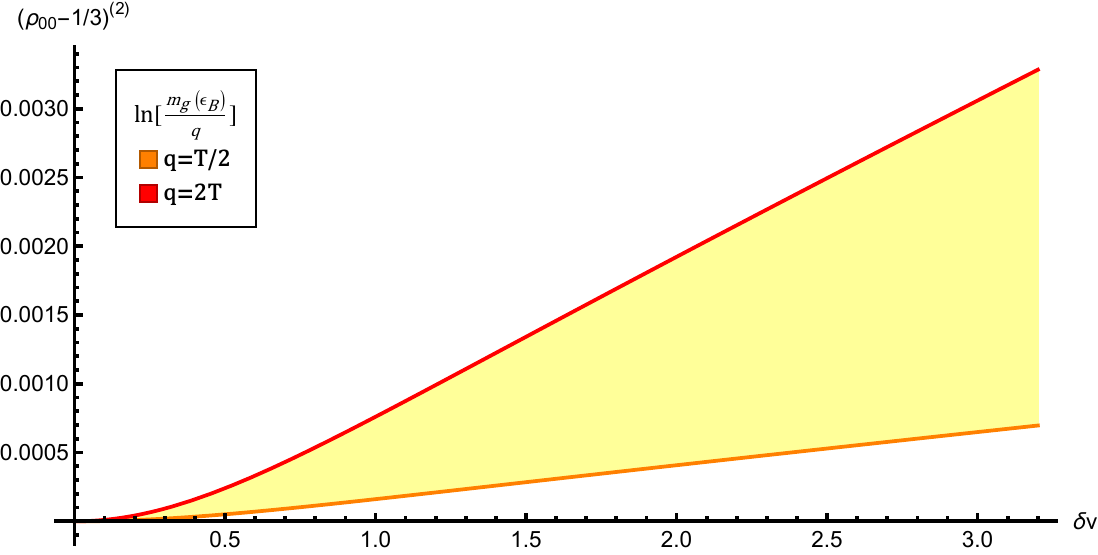}
    \caption{Numerical results of spin alignment up to $\mathcal{O}(\d v^2)$ order of the NLO process. Theory uncertainty is estimated by varying the upper bound of $q$-integration from $T/2$ to $2T$ (see the inset). We take $\a_s=0.3$, binding energy $\e_B=0.052$ GeV and gluon thermal mass $m_\text{g}\simeq1.5853 T$.  \protect}
    \label{NLO-D}
\end{figure}

Now let us turn to numerical simulation for spin alignment of $J/\ps$. Below we work in $J/\ps$'s rest frame. In Bjorken flow with $T=T_0(\tau/\tau_0)^{-1/3}$, the $\tau^{\prime}$ integral provides a time scale relates to temperature. For the parameters, we take initial condition $\tau_0=0.6$ fm/c, $T_0=350$ MeV, and the freeze out temperature $T_c=150$ MeV in this paper. We further perform average in the direction of $J/\ps$ momentum. Consider $J/\ps$ produced at mid-rapidity with $Y=0$, the momentum is perpendicular to beam axis. In Bjorken flow, the momentum distribution is isotropic in transverse plane. The directional average gives $\lag(\d\hat{\bm{v}}\cdot\hat{\bm{l}})^2\rag=1/2$ for quantization axis perpendicular to the beam axis. For the LO process, the numerical results are shown in Fig.~\ref{LOresult}, giving $\rho_{00}>1/3$. The spin alignment results of the NLO process are shown in Fig.~\ref{NLO-D}, with the same sign as the LO process.

\section{Conclusion and Outlook}\label{sec_outlook}

We have calculated polarized dissociation rate of quarkonium spin triplet bound state from spin chromomagnetic coupling in the pNRQCD framework. This is done for the LO gluo-dissociation process and NLO inelastic Coulomb scattering process. The dissociation rate is found to be proportional to fluctuation of the chromomagnetic fields in medium. The spin dependence of the dissociation rate comes from motion of the quarkonium with respect to QGP: in the rest frame of quarkonium, isotropic fluctuation of chrmomagnetic fields in QGP frame becomes anisotropic in quarkonium frame, leading to dependence of the dissociation rate on the spin of quarkonium. The polarized dissociation rate has been expressed as a function of relative velocity between quarkonium and QGP and the quantization axis. Applying the polarized dissociation rate to quarkonium evolution with dissociation effect only in a Bjorken flow, we have found the spin $0$ state to dissociate less than the other spin states, leading to positive $\r_{00}-1/3$.

The NLO polarized dissociation rate is calculated by keeping the logarithmically enhanced terms (either in binding energy or gluon thermal mass) only. It is desirable to perform a more complete analysis beyond the leading logarithmic order, which involves Compton scattering process and interference correction to gluo-dissociation.

A realistic phenomenological analysis of quarkonium evolution is to incorporate regeneration effect. This would require calculation of polarized regeneration rate. Since the regeneration is inverse process of dissociation, it is expected to give a contribution to spin alignment with opposite sign to the dissociation contribution. We leave this for future explorations.

\section*{Acknowledgments}
We thank Yun Guo and Min He for collaboration at early stage of this work. We also thank X.-z. Bai, P. Braun-Munzinger, Y. Jia, D.-f. Hou, X.-j. Yao, K. Zhou, P.-f. Zhuang for stimulating discussions. This work is in part supported by NSFC under Grant Nos 12475148, 12075328.


\newcommand{\setappformulanumber}{%
			\renewcommand{\theequation}{\thesection\arabic{equation}}%
			\setcounter{equation}{0}
}
\appendix

\section{Phase space integration}
\setappformulanumber
\label{sec_app_A}
In this appendix, we evaluate the following integrals for the dissociation rate
\begin{align}\label{four_integrals}
            &\int_Q \text{Im}\[\frac{1}{\left(Q^2-G\right)^2}(3 G-2 m^2)\]\frac{-Q^2}{(Q\cdot n)^2-Q^2}\(\frac{1}{2}+f(q_0)\)\frac{m_Qa^3p_\rel}{(1+a^2p_\rel^2)^3}q^2,\nonumber\\
            &\int_Q \text{Im}\[\frac{1}{\left(Q^2-G\right)^2}\d G^{(2)}\]\(\frac{1}{2}+f(q_0)\)\frac{m_Qa^3p_\rel}{(1+a^2p_\rel^2)^3}q^2,\nonumber\\
            &\int_Q \text{Im}\[\frac{1}{\left(Q^2-G\right)^2}\d G^{(1)}\]\d f^{(1)}\frac{m_Qa^3p_\rel}{(1+a^2p_\rel^2)^3}q^2,\nonumber\\
            &\int_Q \text{Im}\[\frac{1}{Q^2-G}\]\d f^{(2)}\frac{m_Qa^3p_\rel}{(1+a^2p_\rel^2)^3}q^2.
\end{align}
with $p_\rel=(m_Q \left(\text{q0}-\epsilon _B\right))^{1/2}$. The four terms have one-to-one correspondence with those in \eqref{Drr2}. The factor $\frac{a^3p_\rel^2}{(1+a^2p_\rel^2)^3}\frac{m_Q}{p_\rel}$ comes from transition amplitude and the factor $q^2$ comes from the contractions \eqref{contractions_exp}. Since we are concerned with the logarithmically enhanced contributions from $\e_B\sim m_{\text{g}}\ll q(q_0)\ll T$, we may approximate $\frac{1}{2}+f(q_0)\simeq \frac{1}{\b q_0}$ and drop $G$ next to $Q^2$. Denoting $x=\frac{q_0}{q}$, we expect IR divergence at $x\to0$ and collinear divergence at $x\to1$.
We start with the first integral in \eqref{four_integrals}. Using explicit expression of $G$ in \eqref{FG} and simplification above, we find part of the integrand as
\begin{align}
    \text{Im}\[\frac{1}{(Q^2)^2}(3G-2m^2)\]\frac{-Q^2}{(Q\cdot n)^2-Q^2}\frac{1}{\b q_0}\simeq -\frac{\p m_{\text{g}}^2}{q^5\b}.
\end{align}
We have already extract the angle information in the contraction in \eqref{contractions_exp}. The angular integration can then be extracted and performed separately, and a change of variable is used $q_0=x\,q$ to rewrite the integration measure as
\begin{align}
    \int_Q=\int \frac{dq_0q^2dqd\O}{(2\p)^4}=\int q^3dqdx \int \frac{d\O}{(2\p)^4}.
\end{align}
We first perform the $q$-integration from $\e_B/x$ to $\infty$, and then the $x$-integration to obtain
\begin{align}
    \int_{\h}^1 dx\int_{\e_B/x}^\infty dq q^3\(-\frac{\p m_{\text{g}}^2}{q^5\b}\)\frac{m_Qa^3p_\rel}{(1+a^2p_\rel^2)^3}q^2=\frac{3\p^2m_{\text{g}}^2}{16\b}\ln \h.
\end{align}
Note that $x$ has a lower bound $\h=\e_B/q$. Since we are concerned with the coefficient of the logarithm only, we may set $q=T$ to obtain
\begin{align}\label{int0}
    \int q^3dqdx\text{Im}\[\frac{1}{\left(Q^2-G\right)^2}(3 G-2 m^2)\]\frac{-Q^2}{(Q\cdot n)^2-Q^2}\(\frac{1}{2}+f(q_0)\)\frac{m_Qa^3p_\rel}{(1+a^2p_\rel^2)^3}q^2\simeq\frac{3\p^2m_{\text{g}}^2}{16\b}\ln\frac{\e_B}{T}.
\end{align}
To evaluate the second integral in \eqref{four_integrals}, we expand the integrand as
\begin{align}\label{expand}
    \text{Im}\[\frac{1}{(Q^2-G)^2}\d G^{(2)}\]\(\frac{1}{2}+f(q_0)\)\simeq \frac{3 \pi  m_{\text{g}}^2 (\hat{\bm{q}}\cdot \d\bm{v})^2 \left(3 q^2-5 q_0^2\right)}{4 \beta  q^5 \left(q^2-q_0^2\right)}.
\end{align}
The factor $(\hat{\bm{q}}\cdot\d\bm{v})^2$ is taken care of in \eqref{contractions_exp}. We may drop this factor and proceed with the integration as the first one. We find
\begin{align}
    \int q^3dqdx \text{Im}\[\frac{1}{\left(Q^2-G\right)^2}\d G^{(2)}\]\(\frac{1}{2}+f(q_0)\)\frac{m_Qa^3p_\rel}{(1+a^2p_\rel^2)^3}q^2\simeq \frac{3 \pi ^2 m_{\text{g}}^2}{32 \beta}\left(\ln \left(1-x^2\right)\vert_{0}^{1}+3 \ln (x)\vert_{\h}^{1}\right).
\end{align}
The first term in the bracket contains a collinear divergence as $x\to1$. This is regularized by restoring $G$ next to $Q^2$. Note that the $\ln(1-x^2)$ arise from the denominator $q^2-q_0^2$ in \eqref{expand}. With $G$, it is modified into $q^2-q_0^2+m_{\text{g}}^2$ in the collinear limit as $G(q_0=q)=m_{\text{g}}^2$\footnote{One factor of $Q^2-G$ remains in the denominator with the other cancels by $\text{Im}\[\d G^{(2)}\]$.}. This amounts to the modification $\ln(1-x^2)\to\ln(1-x^2+m_{\text{g}}^2/q^2)$
To logarithmic accuracy, we can again set $q=T$ to obtain $\ln(1-x^2)\vert_{0}^{1}\simeq 2\ln\frac{m_{\text{g}}}{T}$. On the other hand, $\ln(x)\vert_{\h}^{1}$ is replaced by in the first integral: $\ln(x)\vert_{\h}^{1}\simeq\ln\frac{\e_B}{T}$. Collecting both IR and collinear logarithms and restoring factors, we have
\begin{align}\label{int1}
    \int q^3dqdx \text{Im}\[\frac{1}{\left(Q^2-G\right)^2}\d G^{(2)}\]\(\frac{1}{2}+f(q_0)\)\frac{m_Qa^3p_\rel}{(1+a^2p_\rel^2)^3}q^2\simeq \frac{3 \pi ^2 m_{\text{g}}^2}{32 \beta}\left(2\ln\frac{m_{\text{g}}}{T} +3 \ln \frac{\e_B}{T}\right).
\end{align}
The third and fourth integrals are evaluated similar to give
\begin{align}
    &\int q^3dqdx \text{Im}\[\frac{1}{\left(Q^2-G\right)^2}\d G^{(1)}\]\d f^{(1)}\frac{m_Qa^3p_\rel}{(1+a^2p_\rel^2)^3}q^2\simeq\frac{\pi ^2 m_{\text{g}}^2}{16 \beta }\left(-\frac{1}{2 x^2}+\ln \left(1-x^2\right)-2 \ln (x)\right)\vert_\h^1,\nonumber\\
    &\int q^3dqdx \text{Im}\[\frac{1}{\left(Q^2-G\right)}\]\d f^{(2)}\frac{m_Qa^3p_\rel}{(1+a^2p_\rel^2)^3}q^2\simeq \frac{\pi ^2 m_{\text{g}}^2 }{32 \beta }\left(\frac{1}{x^2}+\ln \left(1-x^2\right)-2 \ln (x)\right)\vert_{\h}^1.
\end{align}
Although each contribution contains an IR divergence more severe than the logarithmic one, they cancel in the sum. Using the same regularization as the first and second integrals, we obtain
\begin{align}\label{int2}
    &\int q^3dqdx \(\text{Im}\[\frac{1}{\left(Q^2-G\right)^2}\d G^{(1)}\]\d f^{(1)}+\text{Im}\[\frac{1}{\left(Q^2-G\right)}\]\d f^{(2)}\)\frac{m_Qa^3p_\rel}{(1+a^2p_\rel^2)^3}q^2\nonumber\\
    \simeq&\frac{\pi ^2 m_{\text{g}}^2}{32 \beta }\left(3\ln \frac{m_{\text{g}}}{T}-6 \ln \frac{\e_B}{T}\right).
\end{align}



\bibliographystyle{unsrt}\bibliography{spin_chromomagnetic.bib}

\begin{thebibliography}{10}

\bibitem{Liang:2004ph}
Zuo-Tang Liang and Xin-Nian Wang.
\newblock {Globally polarized quark-gluon plasma in non-central A+A
  collisions}.
\newblock {\em Phys. Rev. Lett.}, 94:102301, 2005.
\newblock [Erratum: Phys.Rev.Lett. 96, 039901 (2006)].

\bibitem{Liang:2004xn}
Zuo-Tang Liang and Xin-Nian Wang.
\newblock {Spin alignment of vector mesons in non-central A+A collisions}.
\newblock {\em Phys. Lett. B}, 629:20--26, 2005.

\bibitem{STAR:2017ckg}
L.~Adamczyk et~al.
\newblock {Global $\Lambda$ hyperon polarization in nuclear collisions:
  evidence for the most vortical fluid}.
\newblock {\em Nature}, 548:62--65, 2017.

\bibitem{Becattini:2013fla}
F.~Becattini, V.~Chandra, L.~Del~Zanna, and E.~Grossi.
\newblock {Relativistic distribution function for particles with spin at local
  thermodynamical equilibrium}.
\newblock {\em Annals Phys.}, 338:32--49, 2013.

\bibitem{Fang:2016vpj}
Ren-hong Fang, Long-gang Pang, Qun Wang, and Xin-nian Wang.
\newblock {Polarization of massive fermions in a vortical fluid}.
\newblock {\em Phys. Rev. C}, 94(2):024904, 2016.

\bibitem{Li:2017slc}
Hui Li, Long-Gang Pang, Qun Wang, and Xiao-Liang Xia.
\newblock {Global $\Lambda$ polarization in heavy-ion collisions from a
  transport model}.
\newblock {\em Phys. Rev. C}, 96(5):054908, 2017.

\bibitem{Jiang:2016woz}
Yin Jiang, Zi-Wei Lin, and Jinfeng Liao.
\newblock {Rotating quark-gluon plasma in relativistic heavy ion collisions}.
\newblock {\em Phys. Rev. C}, 94(4):044910, 2016.
\newblock [Erratum: Phys.Rev.C 95, 049904 (2017)].

\bibitem{Wei:2018zfb}
De-Xian Wei, Wei-Tian Deng, and Xu-Guang Huang.
\newblock {Thermal vorticity and spin polarization in heavy-ion collisions}.
\newblock {\em Phys. Rev. C}, 99(1):014905, 2019.

\bibitem{STAR:2022fan}
M.~S. Abdallah et~al.
\newblock {Pattern of global spin alignment of \ensuremath{\phi} and K$^{*0}$
  mesons in heavy-ion collisions}.
\newblock {\em Nature}, 614(7947):244--248, 2023.

\bibitem{ALICE:2022dyy}
Shreyasi Acharya et~al.
\newblock {Measurement of the J/\ensuremath{\psi} Polarization with Respect to
  the Event Plane in Pb-Pb Collisions at the LHC}.
\newblock {\em Phys. Rev. Lett.}, 131(4):042303, 2023.

\bibitem{Pal:2002aw}
Subrata Pal, C.~M. Ko, and Zi-wei Lin.
\newblock {Phi meson production in relativistic heavy ion collisions}.
\newblock {\em Nucl. Phys. A}, 707:525--539, 2002.

\bibitem{Sheng:2022wsy}
Xin-Li Sheng, Lucia Oliva, Zuo-Tang Liang, Qun Wang, and Xin-Nian Wang.
\newblock {Spin Alignment of Vector Mesons in Heavy-Ion Collisions}.
\newblock {\em Phys. Rev. Lett.}, 131(4):042304, 2023.

\bibitem{Sheng:2023urn}
Xin-Li Sheng, Shi Pu, and Qun Wang.
\newblock {Momentum dependence of the spin alignment of the \ensuremath{\phi}
  meson}.
\newblock {\em Phys. Rev. C}, 108(5):054902, 2023.

\bibitem{Xu:2024kdh}
Kun Xu and Mei Huang.
\newblock {Spin alignment of vector mesons induced by local spin density
  fluctuations}.
\newblock {\em Phys. Rev. D}, 110(9):094034, 2024.

\bibitem{Kumar:2023ghs}
Avdhesh Kumar, Berndt M\"uller, and Di-Lun Yang.
\newblock {Spin alignment of vector mesons by glasma fields}.
\newblock {\em Phys. Rev. D}, 108(1):016020, 2023.

\bibitem{Li:2022vmb}
Feng Li and Shuai Y.~F. Liu.
\newblock {Tensor Polarization and Spectral Properties of Vector Meson in QCD
  Medium}.
\newblock 6 2022.

\bibitem{Wagner:2022gza}
David Wagner, Nora Weickgenannt, and Enrico Speranza.
\newblock {Generating tensor polarization from shear stress}.
\newblock {\em Phys. Rev. Res.}, 5(1):013187, 2023.

\bibitem{Sheng:2024kgg}
Xin-Li Sheng, Yan-Qing Zhao, Si-Wen Li, Francesco Becattini, and Defu Hou.
\newblock {Holographic spin alignment for vector mesons}.
\newblock {\em Phys. Rev. D}, 110(5):056047, 2024.

\bibitem{Fu:2023qht}
Baochi Fu, Fei Gao, Yu-Xin Liu, and Huichao Song.
\newblock {The spin alignment of vector mesons with light front quarks}.
\newblock {\em Phys. Lett. B}, 855:138821, 2024.

\bibitem{Zhao:2024ipr}
Yan-Qing Zhao, Xin-Li Sheng, Si-Wen Li, and Defu Hou.
\newblock {Holographic spin alignment of J/\ensuremath{\psi} meson in
  magnetized plasma}.
\newblock {\em JHEP}, 08:070, 2024.

\bibitem{Chen:2024afy}
Jin-Hui Chen, Zuo-Tang Liang, Yu-Gang Ma, Xin-Li Sheng, and Qun Wang.
\newblock {Vector meson\textquoteright{}s spin alignments in high energy
  reactions}.
\newblock {\em Sci. China Phys. Mech. Astron.}, 68(1):211001, 2025.

\bibitem{Braun-Munzinger:2000csl}
P.~Braun-Munzinger and J.~Stachel.
\newblock {(Non)thermal aspects of charmonium production and a new look at J /
  psi suppression}.
\newblock {\em Phys. Lett. B}, 490:196--202, 2000.

\bibitem{Zhao:2007hh}
Xingbo Zhao and Ralf Rapp.
\newblock {Transverse Momentum Spectra of $J/\psi$ in Heavy-Ion Collisions}.
\newblock {\em Phys. Lett. B}, 664:253--257, 2008.

\bibitem{Zhou:2014kka}
Kai Zhou, Nu~Xu, Zhe Xu, and Pengfei Zhuang.
\newblock {Medium effects on charmonium production at ultrarelativistic
  energies available at the CERN Large Hadron Collider}.
\newblock {\em Phys. Rev. C}, 89(5):054911, 2014.

\bibitem{Bodwin:1994jh}
Geoffrey~T. Bodwin, Eric Braaten, and G.~Peter Lepage.
\newblock {Rigorous QCD analysis of inclusive annihilation and production of
  heavy quarkonium}.
\newblock {\em Phys. Rev. D}, 51:1125--1171, 1995.
\newblock [Erratum: Phys.Rev.D 55, 5853 (1997)].

\bibitem{Brambilla:2004jw}
Nora Brambilla, Antonio Pineda, Joan Soto, and Antonio Vairo.
\newblock {Effective Field Theories for Heavy Quarkonium}.
\newblock {\em Rev. Mod. Phys.}, 77:1423, 2005.

\bibitem{Chen:2017jje}
Shile Chen and Min He.
\newblock {Gluo-dissociation of heavy quarkonium in the quark-gluon plasma
  reexamined}.
\newblock {\em Phys. Rev. C}, 96(3):034901, 2017.

\bibitem{Yang:2024ejk}
Di-Lun Yang and Xiaojun Yao.
\newblock {Quarkonium polarization in medium from open quantum systems and
  chromomagnetic correlators}.
\newblock {\em Phys. Rev. D}, 110(7):074037, 2024.

\bibitem{Yan:1980uh}
Tung-Mow Yan.
\newblock {Hadronic Transitions Between Heavy Quark States in Quantum
  Chromodynamics}.
\newblock {\em Phys. Rev. D}, 22:1652, 1980.

\bibitem{Ghiglieri:2020dpq}
Jacopo Ghiglieri, Aleksi Kurkela, Michael Strickland, and Aleksi Vuorinen.
\newblock {Perturbative Thermal QCD: Formalism and Applications}.
\newblock {\em Phys. Rept.}, 880:1--73, 2020.

\bibitem{Bhanot:1979vb}
Gyan Bhanot and Michael~E. Peskin.
\newblock {Short Distance Analysis for Heavy Quark Systems. 2. Applications}.
\newblock {\em Nucl. Phys. B}, 156:391--416, 1979.

\bibitem{Peskin:1979va}
Michael~E. Peskin.
\newblock {Short Distance Analysis for Heavy Quark Systems. 1. Diagrammatics}.
\newblock {\em Nucl. Phys. B}, 156:365--390, 1979.

\bibitem{Song:2005yd}
Taesoo Song and Su~Houng Lee.
\newblock {Quarkonium-hadron interactions in perturbative QCD}.
\newblock {\em Phys. Rev. D}, 72:034002, 2005.

\bibitem{Grandchamp:2001pf}
L.~Grandchamp and R.~Rapp.
\newblock {Thermal versus direct J / Psi production in ultrarelativistic heavy
  ion collisions}.
\newblock {\em Phys. Lett. B}, 523:60--66, 2001.

\bibitem{Grandchamp:2002wp}
L.~Grandchamp and R.~Rapp.
\newblock {Charmonium suppression and regeneration from SPS to RHIC}.
\newblock {\em Nucl. Phys. A}, 709:415--439, 2002.

\bibitem{Chen:2018dqg}
Shile Chen and Min He.
\newblock {Heavy quarkonium dissociation by thermal gluons at next-to-leading
  order in the Quark\textendash{}Gluon Plasma}.
\newblock {\em Phys. Lett. B}, 786:260--267, 2018.

\bibitem{Zhao:2024gxt}
Shouxing Zhao and Min He.
\newblock {Second-order dissociation and transition of heavy quarkonia in the
  quark-gluon plasma}.
\newblock {\em Phys. Rev. D}, 110(7):074040, 2024.

\bibitem{Brambilla:2011sg}
Nora Brambilla, Miguel~Angel Escobedo, Jacopo Ghiglieri, and Antonio Vairo.
\newblock {Thermal width and gluo-dissociation of quarkonium in pNRQCD}.
\newblock {\em JHEP}, 12:116, 2011.

\bibitem{Brambilla:2013dpa}
Nora Brambilla, Miguel~Angel Escobedo, Jacopo Ghiglieri, and Antonio Vairo.
\newblock {Thermal width and quarkonium dissociation by inelastic parton
  scattering}.
\newblock {\em JHEP}, 05:130, 2013.

\bibitem{Brambilla:2010vq}
Nora Brambilla, Miguel~Angel Escobedo, Jacopo Ghiglieri, Joan Soto, and Antonio
  Vairo.
\newblock {Heavy Quarkonium in a weakly-coupled quark-gluon plasma below the
  melting temperature}.
\newblock {\em JHEP}, 09:038, 2010.

\bibitem{Yao:2018sgn}
Xiaojun Yao and Berndt M\"uller.
\newblock {Quarkonium inside the quark-gluon plasma: Diffusion, dissociation,
  recombination, and energy loss}.
\newblock {\em Phys. Rev. D}, 100(1):014008, 2019.

\bibitem{Bellac:2011kqa}
Michel~Le Bellac.
\newblock {\em {Thermal Field Theory}}.
\newblock Cambridge Monographs on Mathematical Physics. Cambridge University
  Press, 3 2011.

\bibitem{Baym:1984np}
G.~Baym.
\newblock {THERMAL EQUILIBRATION IN ULTRARELATIVISTIC HEAVY ION COLLISIONS}.
\newblock {\em Phys. Lett. B}, 138:18--22, 1984.

\bibitem{Jaiswal:2021uvv}
Sunil Jaiswal, Chandrodoy Chattopadhyay, Lipei Du, Ulrich Heinz, and Subrata
  Pal.
\newblock {Nonconformal kinetic theory and hydrodynamics for Bjorken flow}.
\newblock {\em Phys. Rev. C}, 105(2):024911, 2022.

\bibitem{Zhu:2004nw}
Xiang-lei Zhu, Peng-fei Zhuang, and Nu~Xu.
\newblock {J/psi transport in QGP and p(t) distribution at SPS and RHIC}.
\newblock {\em Phys. Lett. B}, 607:107--114, 2005.

\end{thebibliography}

\end{CJK}

\end{document}